\documentclass[fleqn,usenatbib]{mnras}

\usepackage{newtxtext,newtxmath}
\usepackage[T1]{fontenc}
\usepackage{ae,aecompl}

\usepackage{graphicx}	% Including figure files
\usepackage{amsmath}	% Advanced maths commands
\title[SUCCESS]{SoUthern Cluster sCale Extended Source Survey (SUCCESS): A GMRT and MeerKAT study of nine massive galaxy clusters}

\author[Kale et al.]{
R. Kale$^{1}$\thanks{E-mail: ruta@ncra.tifr.res.in},
V. Parekh$^{2,3}$,
M. Rahaman$^{4,5}$,
D. C. Joshi$^{1}$,
T. Venturi$^{6}$,
K. Kolokythas$^{7}$,\newauthor
J. O. Chibueze$^{7,8}$,
S. Sikhosana$^{9,10}$, 
D. Pillay$^{9,10}$
and K. Knowles$^{2}$
\\
$^{1}$National Centre for Radio Astrophysics, Tata Institute of Fundamental Research, S. P. Pune University Campus, Ganeshkhind, Pune 411007, India\\
$^{2}$Centre for Radio Astronomy Techniques and Technologies, Department of Physics and Electronics, Rhodes University, PO Box 94, Makhanda, 6140, South Africa\\
$^{3}$South African Radio Astronomy Observatory (SARAO),  2 Fir Street, Black River Park, Observatory, Cape Town 7925, South Africa \\
$^{4}$ Department of Astronomy, Astrophysics and Space Engineering, Indian Institute of Technology Indore, Indore 453552, India\\
$^{5}$Institute of Astronomy, National Tsing Hua University, Hsinchu, Taiwan\\
$^{6}$INAF, Istituto di Radioastronomia Via Gobetti 101 40129 Bologna, Italy\\
$^{7}$Centre for Space Research, North-West University, Potchefstroom 2520, South Africa\\
$^{8}$Department of Physics and Astronomy, Faculty of Physical Sciences,  University of Nigeria, Carver Building, 1 University Road, Nsukka, Nigeria\\
$^{9}$Astrophysics Research Centre, University of KwaZulu-Natal, Durban 4041, South Africa\\
$^{10}$School of Mathematics, Statistics, and Computer Science, University of KwaZulu-Natal, Westville 3696, South Africa\\
}

\date{Accepted XXX. Received YYY; in original form ZZZ}

\pubyear{2022}

\begin{document}
\label{firstpage}
\pagerange{\pageref{firstpage}--\pageref{lastpage}}
\maketitle

\begin{abstract}
 We aim to carry out a radio study  of the SoUthern Cluster sCale Extended Source Survey (SUCCESS) sample consisting of twenty massive (M$_{500} > 5\times10^{14}$ M$_{\odot}$), nearby (redshift $<0.3$) and southern ($-50^{\circ} < \delta < -30^\circ$) galaxy clusters detected by the Planck satellite and the South Pole Telescope. Here we report targeted GMRT observations (325/610 MHz) for a sub-sample of nine clusters. We also use the first data release of MeerKAT Galaxy Cluster Legacy Survey (1283 MHz) for five of these nine clusters. The properties of the mini-halo in RXC J0528.9-3927, a candidate mini-halo in A3322, the radio halo and candidate double relics in A3399, and the radio halo in RXC J0232.2-4420 are presented. We also report detection of candidate radio relics at distances 1 and 1.9 Mpc from the center of RXC J0232.2-4420. The southeast relic of A3399 is consistent with the radio power - mass scaling relation for radio relics, while the candidate relics around RXC J0232.2-4420 are outliers. This indicates an origin of the candidate relics near RXC J0232.2-4420 to be independent of this cluster and a cluster merger-shock origin for the relic in A3399.  In this sub-sample of clusters 1/9 hosts a radio halo and double relics, 1/9 hosts a radio halo and 2/9 host mini-halos. The dynamical states based on X-ray morphology show that A3399 is a disturbed cluster; however, the radio halo cluster RXC J0232.2-4420 is relaxed, and the mini-halo clusters have intermediate morphologies, adding to the cases of the less commonly found associations.
\end{abstract}


%%%%%%%%%%%%%%%%%%%%%%%%%%%%%%%%%%%%%%%%%%%%%%%%%%
%%%%%%%%%%%%%%%%% BODY OF PAPER %%%%%%%%%%%%%%%%%%

\section{Introduction}
The origin of diffuse, low surface brightness synchrotron sources termed as radio halos, relics and 
mini-halos that 
occur in a fraction of massive clusters is still poorly understood \citep[see][for reviews]
{bru14, wee19review}. 
The radio halos are $\sim$Mpc in extents and occur primarily in the most massive ($\rm{M} > 5\times10^{14} \rm{M}_\odot$) and morphologically disturbed/merging clusters \citep[e.g.,][]{cas10}; though there are a few cases of radio halos reported in relaxed clusters \citep[][]{bon14,som16}, they show signatures of minor mergers as shown in e.g.,  CL1821+643 \citep{kalpar16}.
Radio relics are the tracers of cluster merger shocks  \citep[e.g.,][]{bag06, wee10,kal12,gas14} and have been found in low and high mass ($\rm{M} > 2\times10^{14} \rm{M}_\odot$) clusters \citep[][]{bru14}.
The mini-halos, on the other hand, are found exclusively in cool-core clusters and have been detected in massive systems. In the cool-core sub-sample of the Extended GMRT Radio Halo Survey \citep[EGRHS,][]{ven07,ven08,kal13,kal15} about $50\%$ of the clusters were found to 
host mini-halos \citep{kal15}. Recently it has been shown using a mass-limited ($M > 6\times 10^{14} \rm{M}_\odot$) statistical sample {that $80\%$ of the most massive cool-core clusters  contain mini-halos \citep[][]{gia17}}. The role of feedback from the central galaxy and the sloshing motions within the cool-core are both important in the generation of mini-halos \citep{gia19,laferriere20}.

The short synchrotron cooling time ($\sim0.01-0.1$ Gyr compared to the diffusion time $> 1$ Gyr) of 
radio emitting electrons implies continuous in-situ re-acceleration, likely due to magnetohydrodynamic (MHD) turbulence  in the cases of radio halos \citep[e. g.][]{bru08}  and possibly also in some mini-halos \citep[e. g.][]{gia14,zuh14}.  
The secondary relativistic electrons produced in hadronic collisions are insufficient according to current upper limits based on associated gamma-ray non-detections by {\em Fermi}  \citep[e.g.,][]{bru14}. 

Radio halos with steep synchrotron spectra ($\alpha>1.8$, $S_{\nu} \propto \nu^{-\alpha}$) had been predicted by turbulent re-acceleration models and were discovered in low frequency searches ($<$ GHz) \citep{bru08,mac13,weeren21}.

There have been many radio surveys of galaxy clusters recently, which have resulted in a significant increase in the number of known radio halos, relics, and mini-halos. The Giant Metrewave Radio Telescope (GMRT) was used for the first major low-frequency surveys of clusters.
One of the largest radio surveys of galaxy clusters was the Extended GMRT Radio Halo Survey \citep[EGRHS,][]{ven07,ven08,kal13,kal15}. 
EGRHS probed the most X-ray luminous ($L_X > 5 \times 10^{44}$ erg s$^{-1}$) clusters in the redshift range 0.2 - 0.4 at 610 MHz 
with the GMRT and provided new detections of radio halos and mini-halos as well
as the first upper limits \citep[][]{ven08,bru07,bru09}, however it was limited to declinations, $\delta>-31^{\circ}$.

Recent surveys have focused on nearly mass-complete samples extracted from the detections of clusters using the Sunyaev Zeldovich effect such as the Planck satellite \citep[e. g.,][]{2016A&A...594A..27P}. A radio survey of 75 clusters with $\delta > - 40^{\circ}$, in the redshift range 0.03 - 0.33 and with M$_{500} \gtrsim 6\times10^{14}$ M$_{\odot}$ was carried out by \citet{cuc15,cuciti21a,cuciti21b}. 
Using LOFAR (Low Frequency ARray, \citet{2013A&A...556A...2V}) observations at 144 MHz,  \citet{2020NatAs.tmp..226D} found 9 new radio halos from a sample of 20 clusters.

\begin{table*}
    \centering
    \caption{SUCCESS sample properties. Columns are: (1) Cluster name, (2) Alternative names, (3) redshift \citep{2016A&A...594A..27P}, (4) Mass \citep{2016A&A...594A..27P}, (5) X-ray luminosity \citep[][]{2011A&A...534A.109P}, and (6) Morphological type \citep{2017ApJ...846...51L}. {The radio observations of nine clusters in the first sector are presented in this work.}}
    \resizebox{\textwidth}{!}{
    \begin{tabular}{llllllllll}
    \hline
    \hline
    (1)&(2)&(3)&(4)&(5)&(6)\\
    Name   &Alternative Names &z  &M$_{500}$      &L$_{X}$     &Type   \\
            & &  &$10^{14} \mathrm{M}_{\odot}$ &$10^{44}$ erg s$^{-1}$  \\
            \hline
       RXC J1358.9-4750    & PSZ1 G314.46+13.54&0.07 &$5.285^{+0.255}_{-0.236}$  &2.45  &\\
         {RXC J0232.2-4420}       &{PSZ1G260.00-63.45} &0.28 &$7.541^{+0.334}_{-0.323}$  &8.50 &Relaxed\\
        PSZ2 G313.84+19.21     &PSZ1 G313.85+19.21&0.28&$6.694^{+0.539}_{-0.554}$&&\\
         {RXC J0528.9-3927}   &{PSZ1 G244.35-32.15}, RBS0653 &0.28 &$7.406^{+0.369}_{-0.389}$ &14.5 & Mixed\\
           A3396 &SPT-CL J0628-4143 &0.17&$3.842^{+0.365}_{-0.424}$ & 4.33 &\\%4.230388
           A3322 & PSZ1 G250.92-36.24, RXC J0510.2-4519&0.20 &$5.737^{+0.290}_{-0.299}$  &4.96 &Mixed\\
           A3399 &SPT-CL J0637-4829, RXC J0637.3-4828 &0.20  &$5.306^{+0.272}_{-0.272}$ &3.46  &Disturbed\\
            A3343& RXC J0525.8-4715, SPT-CL J0525-4715&0.19 &$4.670^{+0.313}_{-0.324}$ &3.59 &Relaxed\\
            A3937&PSZ2G345.32-59.97, SPT-CL J2254-4620 &0.27 &$5.093^{+0.474}_{-0.489}$ & &  \\
    \hline
  A3378 & PSZ1 G241.76-24.01, RXC J0605.8-3518 &0.14 &$5.588^{+0.198}_{-0.221}$ &4.23 & Relaxed \\%No
    A3364 &PSZ1 G236.93-26.65, RXC J0547.6-3152 &0.15 & $4.889^{+0.292}_{-0.320}$&4.32 & Relaxed\\ %No
    A3888 &PSZ1 G003.93-59.42, RXC J2234.5-3744 & 0.15&$7.190^{+0.264}_{-0.259}$ & 6.38& Mixed\\ %RH 2016MNRAS.459.2525S
    A3016 &PSZ1 G256.55-65.69, RXC J0225.9-4154 &0.22 &$6.082^{+0.282}_{-0.329}$ &6.07 & \\%No
    A3984 & PSZ1 G358.96-67.26, RXC J2315.7-3746 &0.18 &$5.225^{+0.317}_{-0.302}$ &4.57 & \\%No
    RXC J1317.1-3821 &PSZ1 G308.44+24.24 & 0.26&$6.286^{+0.548}_{-0.551}$ & 7.15& \\%No2015MNRAS.453.1201H BCG
    PSZ2 G348.43-25.50 & PSZ1 G348.43-25.50&0.25 &$5.813^{+0.493}_{-0.542}$ & & \\%No
    RXC J0532.9-3701 & PSZ1 G241.75-30.89&0.27&$7.005^{+0.335}_{-0.345}$ &6.20 & Relaxed\\%No
    SPT-CL J0051-4834	&PSZ2 G303.03-68.49 & 0.19& $3.214^{+0.425}_{-0.513}$&2.04 & \\%No
    SPT-CL J2211-4833	&PSZ2 G347.27-52.46 &0.24 &$3.422^{+0.478}_{-0.515}$ & & \\%No
    A3783	&RXC J2134.0-4240, SPT-CL J2134-4238 &0.20 &$4.207^{+0.384}_{-0.396}$ & & \\%No
    \hline
    \end{tabular}
    }
    \label{clusinfo}
\end{table*}

Clusters in the southern hemisphere and especially at declinations, 
$\delta<-31^{\circ}$ have remained outside the earlier cluster samples due to the increasing asymmetry of the beams and limited reach to low declinations of the northern telescopes such as the GMRT and the VLA. 
A handful of radio halos and relics sources were known in the southern sky with the observations using the available telescopes \citep[e. g.][]{2000ApJ...544..686L,1997MNRAS.290..577R}. The recent radio surveys with the MeerKAT and the Murchison Widefield Array have boosted the discoveries of diffuse radio emission in the clusters in the southern sky \citep[e. g.][]{2021PASA...38...10D, 2021MNRAS.504.1749K, 2022A&A...657A..56K}. 
In this paper, we present radio observations towards a sample of nine nearby southern galaxy clusters with the aim of searching for diffuse radio emission. For five of these clusters, we also present the MeerKAT images obtained from the recently public archive (MeerKAT Cluster Legacy Survey).

The paper is organised as follows: the sample is presented in Sec.~\ref{simsample}. GMRT observations and data reduction are described in Sec.~\ref{sec:obsn}. The MeerKAT GCLS sample and their products that are used in this study are briefly
described in  Sec.~\ref{meerobs}, with the results of the current study of 9 clusters described in Sec.~\ref{results}. In Sec.~\ref{discussion}, the results are discussed in comparison to similar literature studies, based on their radio morphological occurrence, radio power, and mass. Lastly, the conclusions are presented in Sec.~\ref{conclusions}.

We have used $\Lambda$CDM cosmology with $\mathrm{H}_{0} = 71$ km s$^{-1}$ Mpc$^{-1}$, 
$\Omega_\mathrm{M} = 0.27$ and $\Omega_\Lambda =0.73$ in this work.

\section{Simulation of radio observations and sample selection}\label{simsample}
Imaging diffuse extended sources with structures ranging from small to large scales is a challenge for radio interferometers. The crucial requirement is to have uv-coverage that is good enough in the relevant angular scales for the source and, if possible, even beyond. We have investigated the effects of uv-coverage on the imaging of extended radio sources such as radio halos for the GMRT and the uGMRT \citep{2017ExA....44..165D}. The GMRT is located at latitude $19^{\circ}05'47''$N, and the limiting elevation for the antennas permits observations of the sky up to the declination of $-53^{\circ}$. 
We simulated radio halos (1 Mpc) with typical surface 
brightnesses at 610 and 330 MHz located at $\delta = -50^{\circ}$ and made simulated observations of these. The visibilities were then cleaned (CASA task `clean'), and the recovery of the emission as compared to that in the model was checked. It was found that radio halos of sizes half of the largest angular scale implied by the shortest baseline can be recovered above $90\%$. 
For southern declinations, the short baseline 
coverage improves (projected baseline lengths are shorter) 
and Mpc-sized radio halos in clusters at redshifts 0.03 onwards can be recovered in 325 MHz observations. Mpc-sized radio halos in clusters at redshifts $>0.15$ can be fully  ($>90\%$) recovered at 610 MHz with the GMRT.

\subsection{SUCCESS Sample}\label{sample}
Based on the simulations described above, {we selected a sample of southern galaxy clusters to search for diffuse radio emission with the GMRT}. Given the scaling relation between the radio power of radio halos and cluster mass, we aimed to search for radio halos in massive clusters which were not yet part of the earlier samples due to the declination cut in their selection.
We selected galaxy clusters from the Planck \citep{PC2015} and the South Pole Telescope (SPT) catalogues \citep{SPT2015}, that satisfied the following criteria:\\
 I) $z<0.3$, \\ 
 II)  $\mathrm{M}_{500} > 5\times10^{14}\mathrm{M}_\odot$ and \\
III) $-50^{\circ}<\delta<-30^{\circ}$.\\
{We further limited the sample from the Planck to clusters with spectroscopic redshifts.}
This resulted in a sample of 13 clusters from the Planck and 9 clusters from the SPT. 
The clusters, RXC J0232.2-4420 and A3322 appeared in both these sub-samples and thus the total sample consists of 20 clusters (Table~\ref{clusinfo}). The redshifts of the sample clusters are between 0.07 and 0.28. The masses of the 
clusters are in the range {3.2} - 7.5 $\times10^{14}$ M$_{\odot}$. The updated masses from \citet{2016A&A...594A..27P} are reported here and thus are a bit different from those 
used for the cluster selection ($\mathrm{M}_{500} > 5\times10^{14}\mathrm{M}_\odot$). The morphological types of the cluster as listed by \citet{2017ApJ...846...51L} are provided where available. 
When the sample was undertaken for radio study, A3888 was the only 
cluster in the sample where a radio halo was known \citep{2016MNRAS.459.2525S}. In this paper, we focus on nine clusters from this sample that we observed with the GMRT. Among these nine, A3396 and A3343 in particular are the clusters with masses $3.8$ and $4.7 \times10^{14} \rm{M}_\odot$ that are lower than the threshold (Table~\ref{clusinfo}), due to the revised value after our
original selection.

\section{GMRT Observations and data reduction}\label{sec:obsn}
The nine clusters were observed with the GMRT under the proposal codes 31\_065 and 32\_043 (PI R. Kale). The details of observations are given in Table ~\ref{obsn}.
The GMRT Software Backend was used to record data using 32 MHz bandwidth spread over 256 channels.
Depending on the availability of the source in the sky and the simulation predictions, the clusters were observed for 5 - 7 hr duration as given in Table ~\ref{obsn}. 
We used the data analysis pipeline CAsa Pipeline-cum-Toolkit for Upgraded GMRT data reduction - CAPTURE \citep{2021ExA....51...95K}. The standard steps of flagging, calibration, and imaging are followed in the CAPTURE. Self-calibration with phase-only and finally with amplitude and phase solutions was carried out in 6 - 8 iterations. The imaging was carried out with a choice of robust$ = 0$ (Briggs' weighting scheme, \citet{1995PhDT.......238B}). 
The flux scale of Perley-Butler 2017 was used for absolute flux calibration. The summary of the rms noise obtained in the robust$ = 0$ images and the corresponding beams are reported in Table~\ref{obsn}. Two GMRT datasets were not usable due to images with heavily compromised sensitivity (poor calibration, loss of several antennas to power failure, or other issues).

We used the "Gaussfit'' option in CASA viewer to obtain the flux densities of discrete sources in the images. The error on the fit was added in quadrature with the absolute flux density error ($\sigma_{\rm abs}$) to obtain the reported error on the flux density. We used $\sigma_{rms} = 0.1$ at 610 and 325 MHz. 
The method for separating the diffuse and compact emission in cases of suspected diffuse sources is described in the individual subsections below. The error on the flux density of extended sources was calculated according to $\sqrt{(\sigma \sqrt N_{\mathrm b})^2 + (\sigma_{\mathrm{abs}} S_{610\mathrm{MHz}})^2}$,
where $N_{\mathrm b}$ is the number of beams in the extent of the emission, and $\sigma_{\rm abs}$ is the percentage error on the absolute flux density scale. 

\section{MeerKAT} \label{meerobs}
MeerKAT is a next-generation interferometer that comprises 64 antennas and one of the precursors of the Square Kilometer Array, located in the Karoo desert, South Africa \citep{2018ApJ...856..180C,2009IEEEP..97.1522J}. 
In this paper, we made use of the data from the MeerKAT Galaxy Cluster Legacy Survey (MGCLS), which began a program of long-track observations of galaxy clusters in the southern sky \citep{2022A&A...657A..56K}. This project consists of observations of 115 galaxy clusters at $L$-band with a total observation time of $\sim$ 1000 hours. The continuum observation was made with full polarisation, 4096 channels, and 8sec integration time, which resulted in images with sub-10$''$ resolution and less than 5 $\mu$Jy beam$^{-1}$ rms sensitivity. All MGCLS data were calibrated and imaged with the procedure described in \citep{2020ApJ...888...61M}. The calibration and imaging steps were performed using the Obit package\footnote{\url{https://www.cv.nrao.edu/$\sim$bcotton/Obit.html}} \citep{2008PASP..120..439C}. For more details about the MGCLS observations, data analysis, and image processing, {we refer readers to check \citet{2022A&A...657A..56K}, with the full MGCLS catalog of radio images and their properties to be shown in a follow-up study \citet[][Kolokythas et al. (in prep)]{}.}

From the MGCLS catalogs, we found observations of six SUCCESS clusters- RXC J1358.9-4750, RXC J0232.2-4420, RXC J0528.9-3927, A3399, A3322, and A3343. {We used the data products of five clusters (except RXC J1358.9-4750) available from the MGCLS survey. The image of RXC J1358.9-4750 was not used due to compromised image quality, likely due to calibration errors.} In particular the primary beam corrected total intensity images at full resolutions $\sim 7''$, and those convolved to $15''$ were used. The rms for the full resolution images is in the range $2 - 7 \mu$Jy beam$^{-1}$ and that for the images with $15''$ resolutions is $7 - 15 \mu$Jy beam$^{-1}$. 
The MeerKAT observations are summarised in Table~\ref{obsn}. We used PyBDSF \citep{2015ascl.soft02007M} for finding the discrete radio sources in the images.

\begin{table*}
\caption{Radio observation summary for the SUCCESS sample. Columns are: (1) Cluster name, (2) Right Ascension, (3) Declination, (4) Date of observation, (5) Telescope, (6) Observing frequency and duration, (7) root mean square noise in the image with robust $= 0.0$, (8) Beam and position angle (p. a.), and (9) Remark. The remarks regarding detection of diffuse radio emission (RH $=$ radio halo, MH $=$ Mini-halo candidate, DR $=$ double relic, c$=$ candidate), previous reference (if any), and comments on the data quality are provided as applicable (x = We have not used this data due to the poor SNR of the data).  \label{obsn}}
\resizebox{\textwidth}{!}{
\begin{tabular}{lllllllll}
\hline
\hline
\noalign{\smallskip}
    (1)&(2)&(3)&(4)&(5)&(6)&(7)&(8)&(9)\\
Name              &RA$_{J2000}$         &DEC$_{J2000}$        &Date    & Tele-                   &$\nu$, t$_{obs}$ & rms & beam & Remark\\
                 & hh mm ss     & $^{\circ}$ $'$ $''$          &                  &scope &MHz, hr & mJy beam$^{-1}$ &$''\times''$, p. a.&\\
\hline
\noalign{\smallskip}
RXC J1358.9-4750           &13 58 42.64 &-47 48 41.9 &24 Oct 2016 &GMRT&325, 5&0.50&$19.4\times6.7, 6.4^{\circ}$&\\
RXC J0232.2-4420       &02 32 18.7 &-44 20 41.0 &18 Feb. 2017&GMRT &610, 6&0.04&$9.2\times3.9, 6.3^{\circ}$&RH$^{\ddag}$ \\
       & & & 13 Jan 2019&MeerKAT &1283, 9&0.0026&$7.1\times7.1$, $86.5^{\circ}$&RH$^{\dag}$\\
       & & &            &MeerKAT &1283, 9& 0.01 & $15''\times15''$& \\
PSZ1 G313.85+19.21             &13 47 35.80 &-42 28 02.6 & 08 Jan. 2017&GMRT &610, 6&0.065&$10.7\times4.0, 14.7^{\circ}$&\\
{RXC J0528.9-3927}$^\dag$      &05 28 48.82 &-39 27 40.0 &13 Dec. 2016&GMRT &610, 7&0.035&$7.5\times4.6, 15.9^{\circ}$ &{MH}\\
       & & & 21 Jan 2019&MeerKAT&1283, 10&0.0026&$7.3\times7.3$,$31.7^{\circ}$&MH$^{\dag}$\\
              & & &            &MeerKAT &1283, 10& 0.0116 & $15''\times15''$& \\
A3396                &06 28 49.2  &-41 43 30   &19 Feb. 2017 &GMRT&325, 6&1.5&$14.4\times7.5, 6.3^{\circ}$&\\
{A3322}             &05 10 26.66 &-45 21 02.3 &15 Aug. 2017 &GMRT&610, 5 &0.05&$9.6\times4.1, 14.8^{\circ}$&{MHc}\\
                          &  & & 12 Jan 2019&MeerKAT &1283,8 & 0.0069 & $7.6\times7.5$, $24.2^{\circ}$ & {MHc$^{\dag}$} \\
{A3399}                &06 37 23.2  &-48 29 16   &22 May 2017&GMRT&610, 5 &-&-&x\\
       & & &17 Feb 2019&MeerKAT &1283, 12&0.0029&$7.5\times7.3$,$-83.9^{\circ}$& {RH, DRc$^{\dag}$}\\
              & & &            &MeerKAT &1283, 12& 0.015 & $15''\times15''$& \\
A3343                &05 25 49.3  &-47 15 20   &22 Jul. 2017 &GMRT &610, 5 &-&-&x\\%No GMRT
       & & & 11 Jan 2019&MeerKAT&1283, 8&0.003&$8.1\times7.7$,$41.8^{\circ}$\\
              & & &            &MeerKAT &1283, 8& 0.007 & $15''\times15''$& Not shown\\
A3937                &22 54 21.1  &-46 20 37   &20 Jul. 2017 &GMRT&610, 5&0.07&$8.6\times3.6, 4.9^{\circ}$&\\%Pulsar obsn\\
\hline \noalign{}
%\hline
%\hline 
\end{tabular}
}
${\ddag}$ \citep{2019MNRAS.486L..80K}; 
${\dag}$ \citep{2022A&A...657A..56K} 
%\\
\end{table*}

\begin{figure*}
    \centering
     \includegraphics[height = 6.5cm]{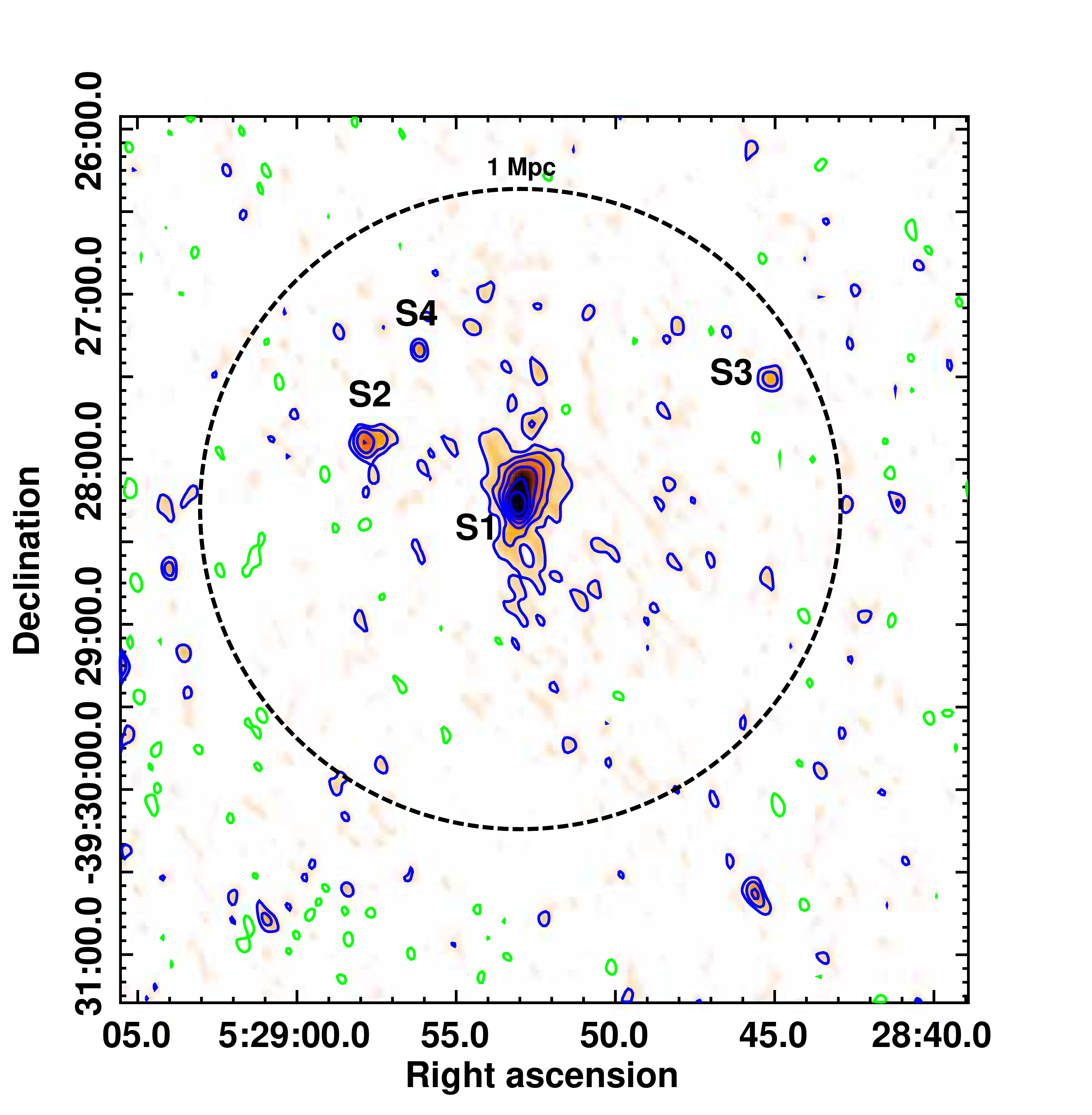}
    \includegraphics[height = 6.5cm]{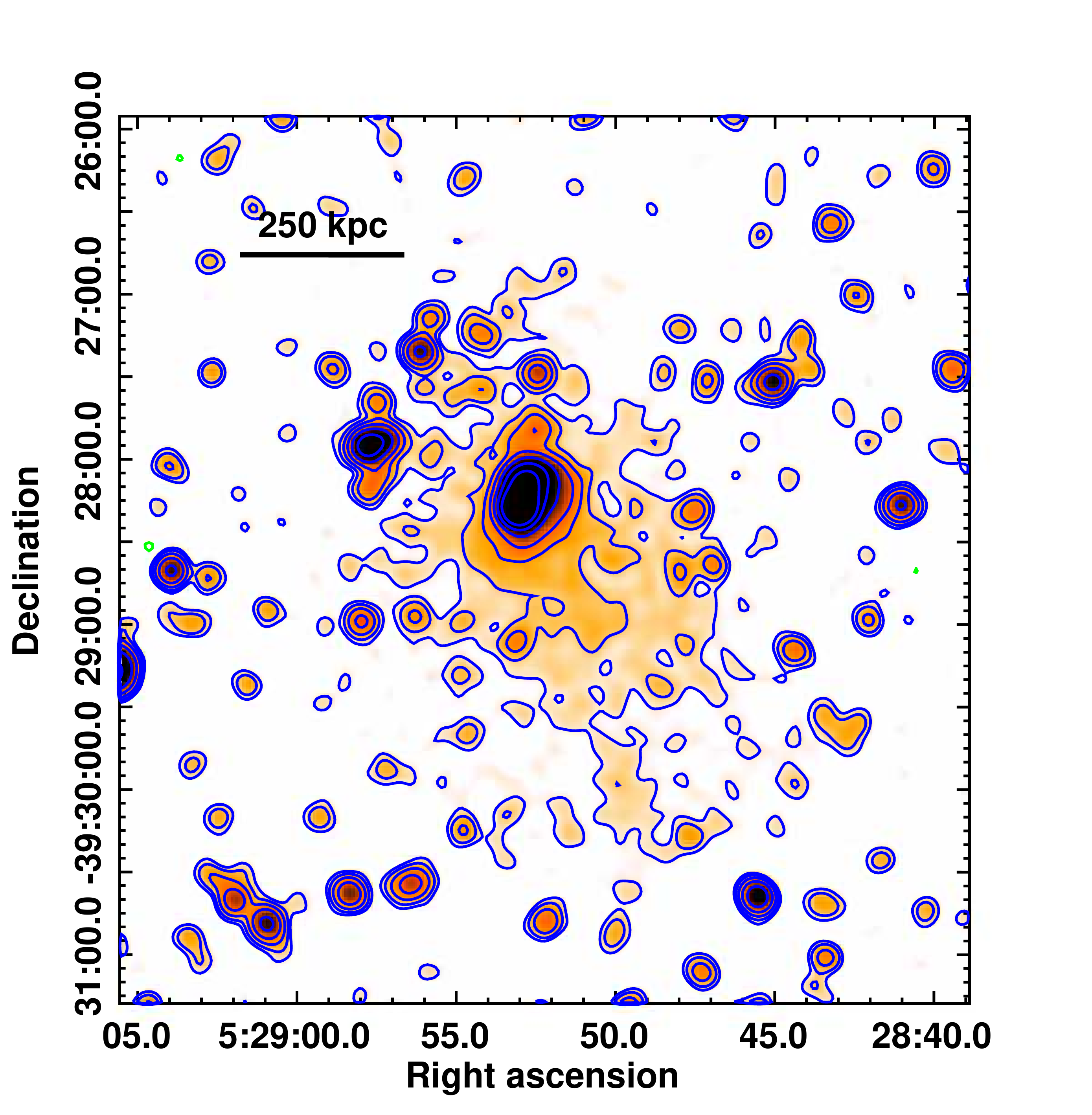}\\
    \includegraphics[height = 6.5cm]{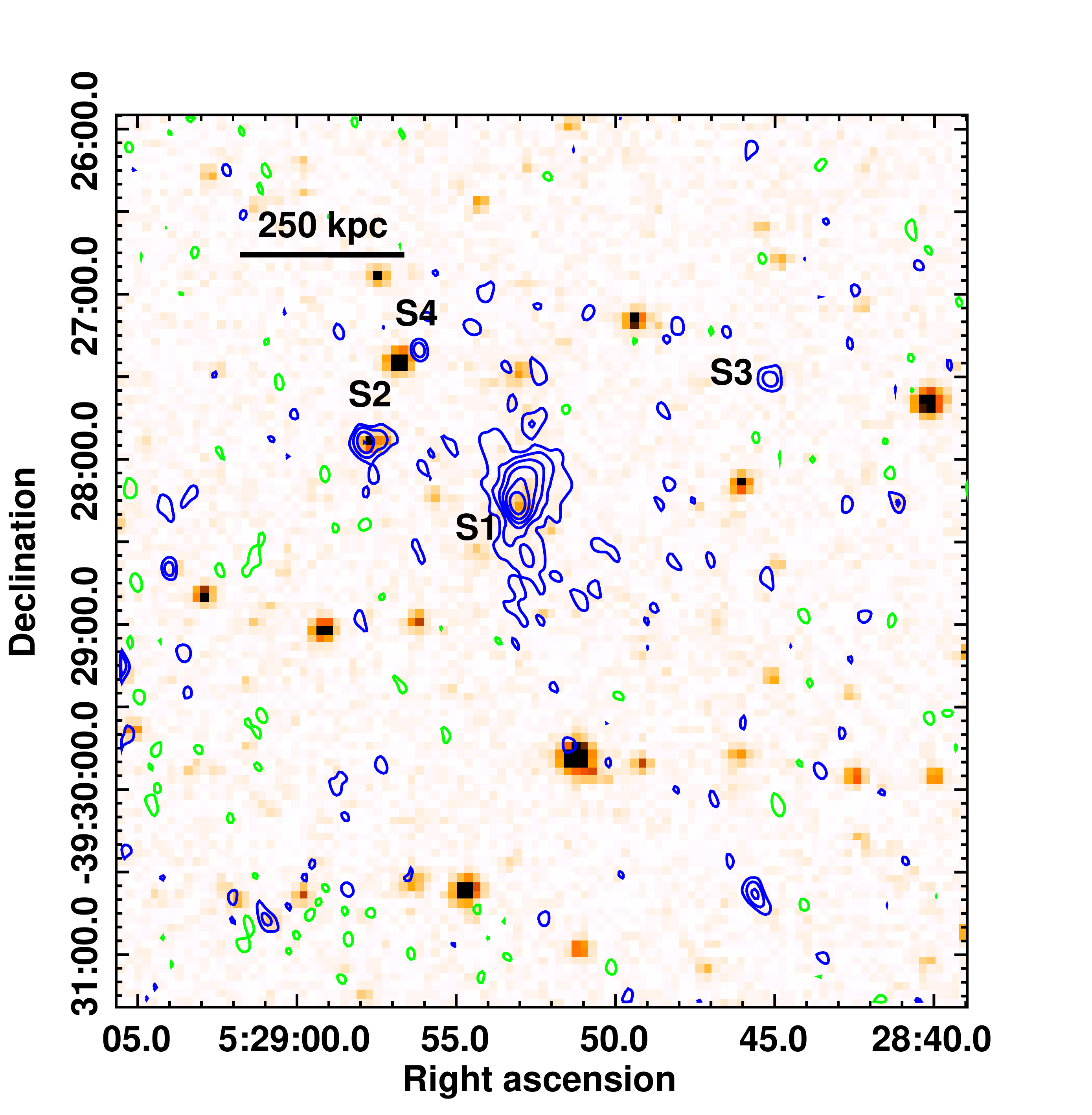}
    \includegraphics[height=6.5cm]{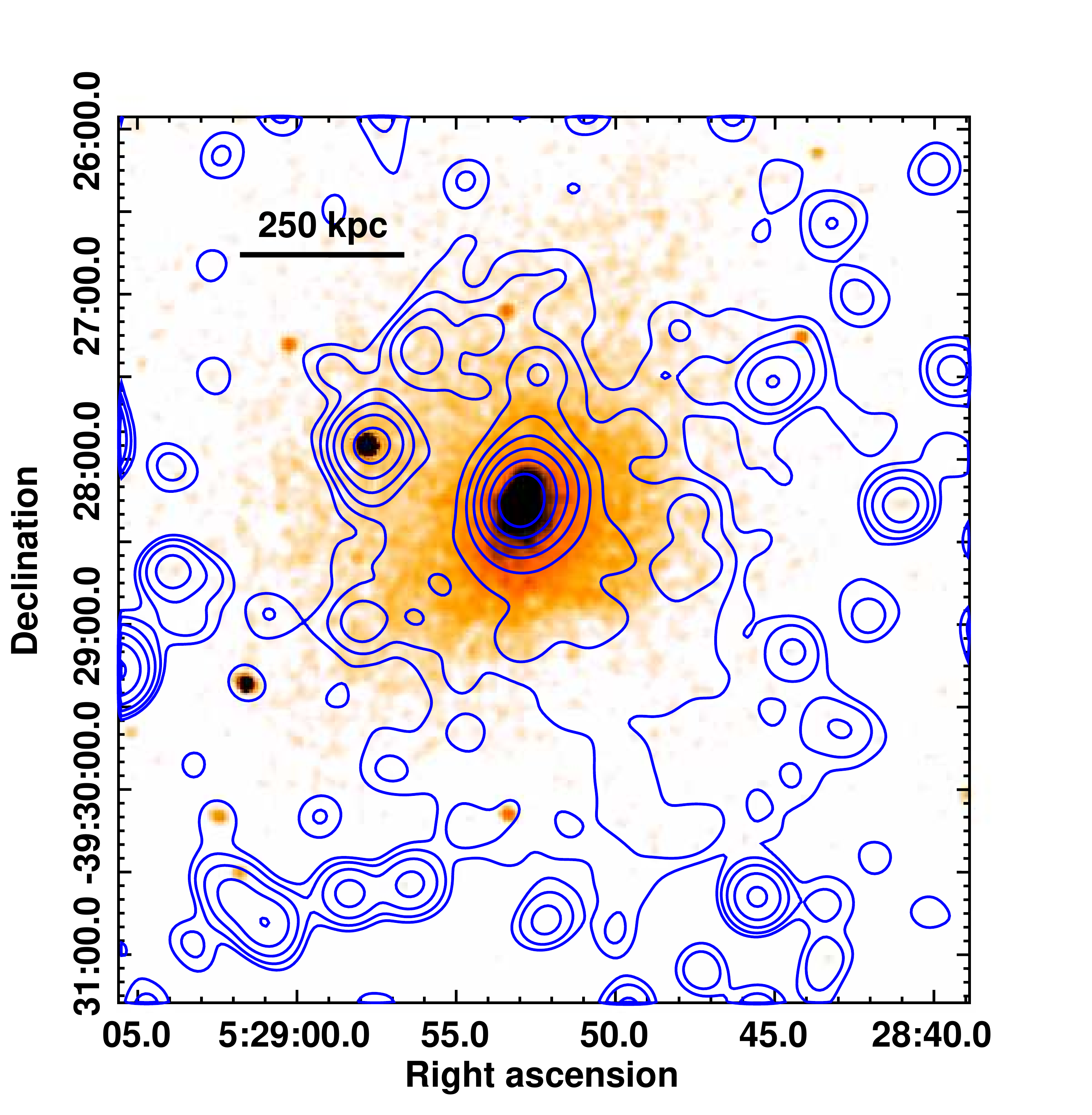}\\
    \includegraphics[height = 6.5cm]{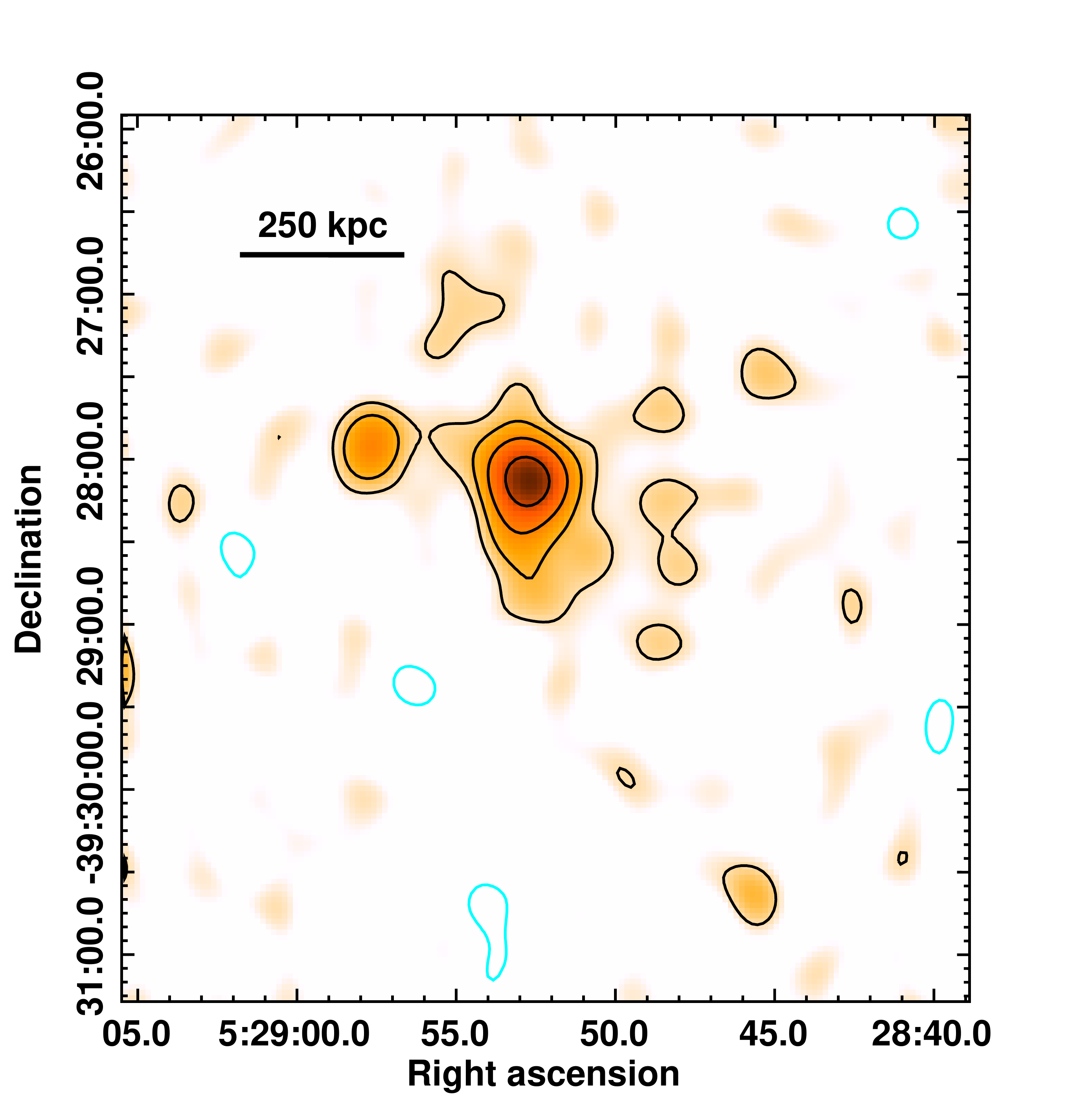}
    \includegraphics[height=6.5cm]{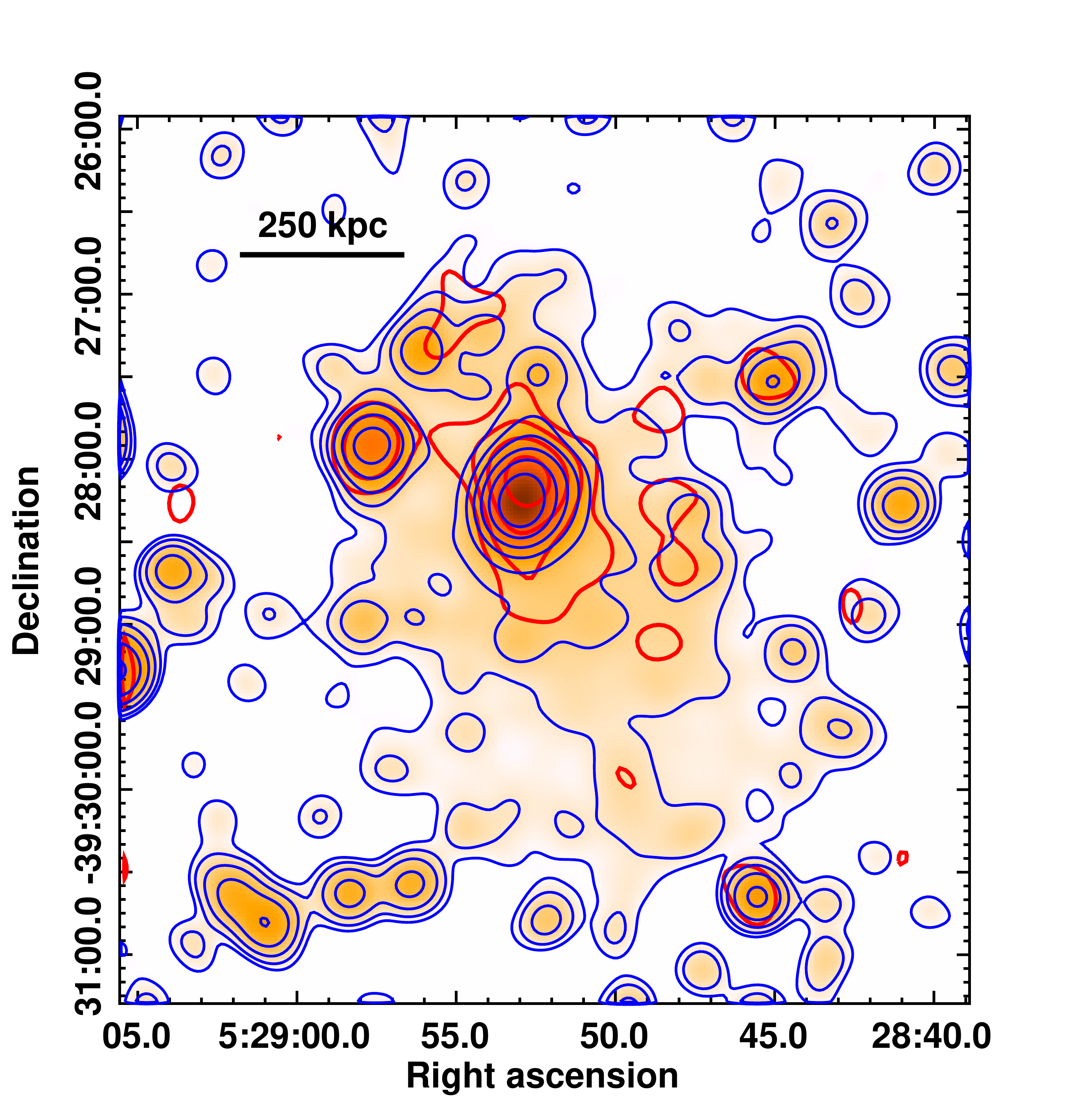}
    \caption{\textbf{RXC J0528.9-3927}: Top left: GMRT 610 MHz image is shown in colour and contours. Contour levels are at $[-1, 1, 2, 4,...]\times3\sigma_{\mathrm{rms}}$ where $\sigma_{\mathrm{rms}} = 0.035 $mJy beam$^{-1}$.
    Positive contours are in blue and negative in green. A circle of diameter 1 Mpc is shown. Top right: MeerKAT high-resolution image (Table ~\ref{obsn}) shown in contours and colour. The contour levels are at $[-1, 1, 2, 4,...]\times3\sigma_{\mathrm{rms}}$ where $\sigma_{\mathrm{rms}} = 0.0026 $ mJy beam$^{-1}$.
    Centre left: The contours from the top left panel are overlaid on the DSS2-red image shown in colour. Centre right: Chandra X-ray image shown in colour overlaid with blue contours of MeerKAT $15''\times15''$ resolution image. The blue contour levels are at $[-1, 1, 2, 4,...]\times3\sigma_{\mathrm{rms}}$ where $\sigma_{\mathrm{rms}} =  0.0046$ mJy beam$^{-1}$.
    Bottom left: GMRT 610 MHz image after subtracting discrete sources. 
    The contours are at $[-1, 1, 2, 4,...]\times3\sigma_{\mathrm{rms}}$ where $\sigma_{\mathrm{rms}} = 0.116$ mJy beam$^{-1}$.
    Black contours are positive, and cyan contours are negative. The beam size is $20.4''\times17.3''$, p. a. 23.5$^{\circ}$. Bottom right: MeerKAT $15''\times15''$ resolution image shown in blue contours and in colour scale. The red contours are the positive contour levels from the GMRT image in the bottom left panel. The blue contour levels are the same as in the center-right panel.
    }
    \label{G244}
\end{figure*}

\begin{figure*}
\centering
\begin{tabular}{lccr}
\includegraphics[width=\columnwidth]{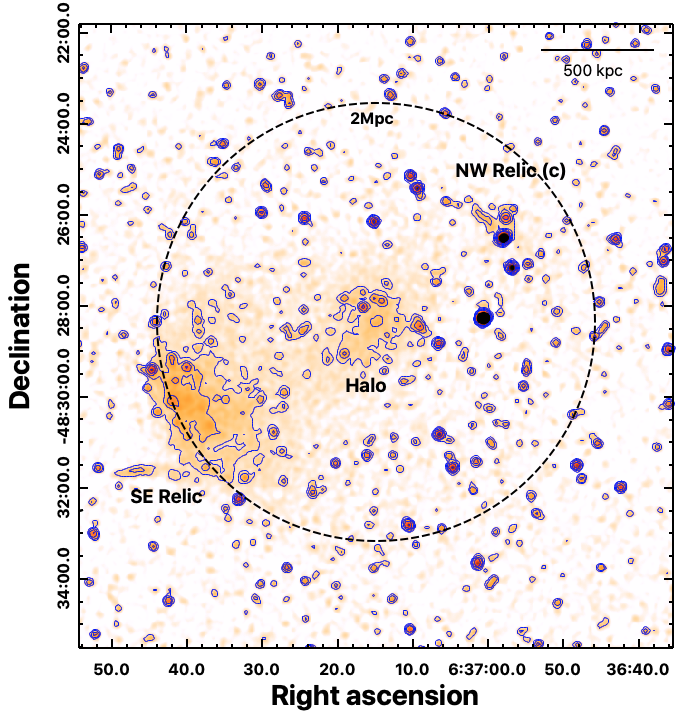} &
\includegraphics[width=\columnwidth]{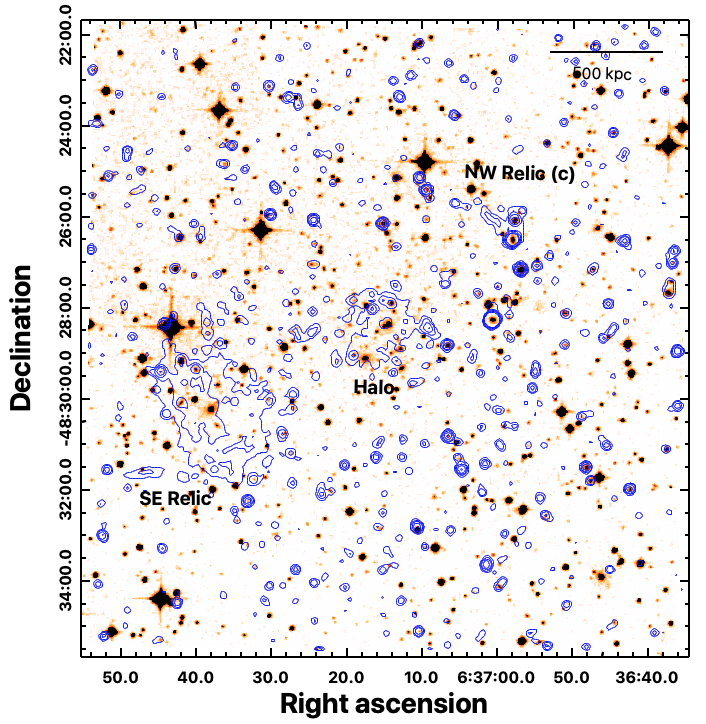} \\
\includegraphics[width=\columnwidth]{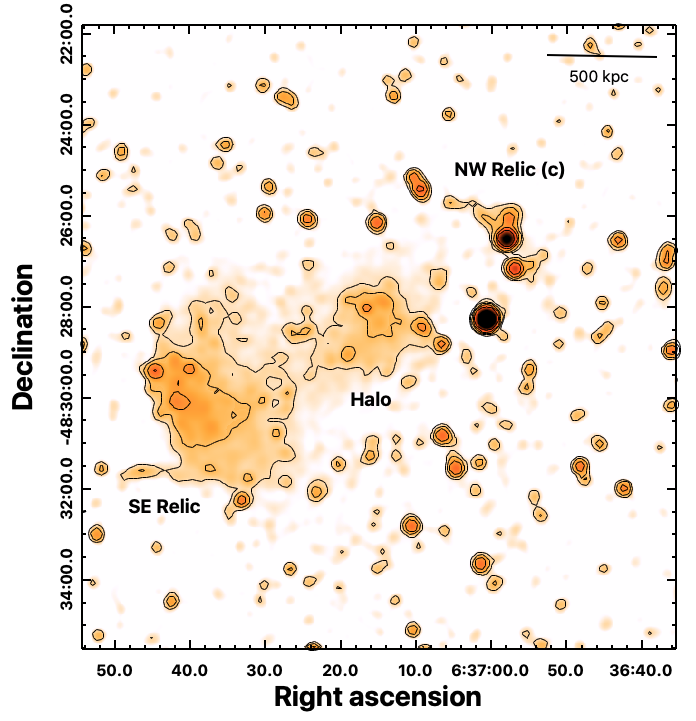} & \includegraphics[width=\columnwidth]{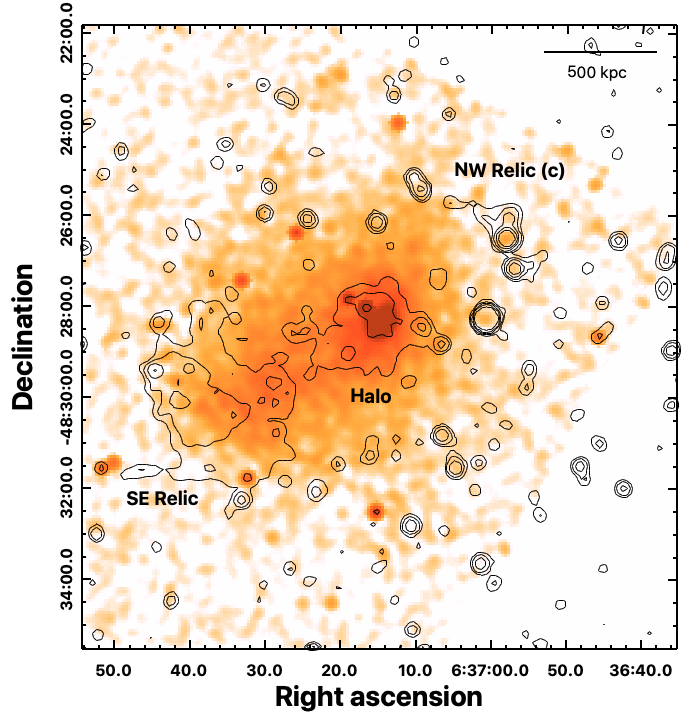} \\
\end{tabular}
\caption{\textbf{A3399}: Top left: MeerKAT's native 7$\arcsec$ resolution 1283 MHz radio image is shown in colour and contours (positive contour in blue and negative contours are in green). Contour levels are drawn at [-1, 1, 2, 4, 8, 16, ...]$\times 5 \sigma_{\mathrm{rms}}$, where $\sigma_{\mathrm{rms}} = 2.9 \mu$Jy beam$^{-1}$. 
A black dashed circle of 2 Mpc diameter is shown, which is centered at the X-ray peak. Top right: The r band DSS2 optical image cutout, overlaid with contours from the top left panel. Bottom left: Low resolution (15$\arcsec$ beam) 1283 MHz MeerKAT image is shown in colour and contours (positive contours in black and negative contours are in green). Contour levels are drawn at [-1, 1, 2, 4, 8, 16, ...]$\times 3  \sigma_{\mathrm{rms}}$, where $ \sigma_{\mathrm{rms}} = 15 \mu$ Jy beam$^{-1}$. 
Bottom right: \textit{Chandra} X-ray surface brightness map overlaid with low-resolution MeerKAT radio contours is shown. Contour levels are the same as the bottom left panel. 
} 
\label{fig:J0637}
\end{figure*}

\begin{figure*}
    \centering
            \includegraphics[height = 6.5cm]{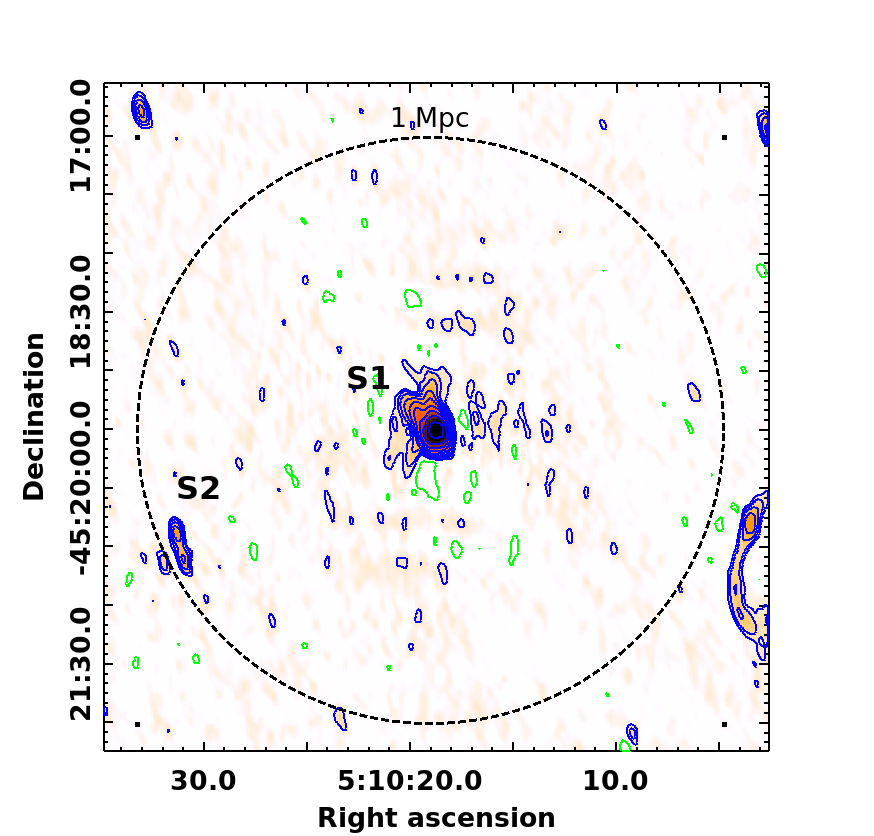}
    \includegraphics[height = 6.5cm]{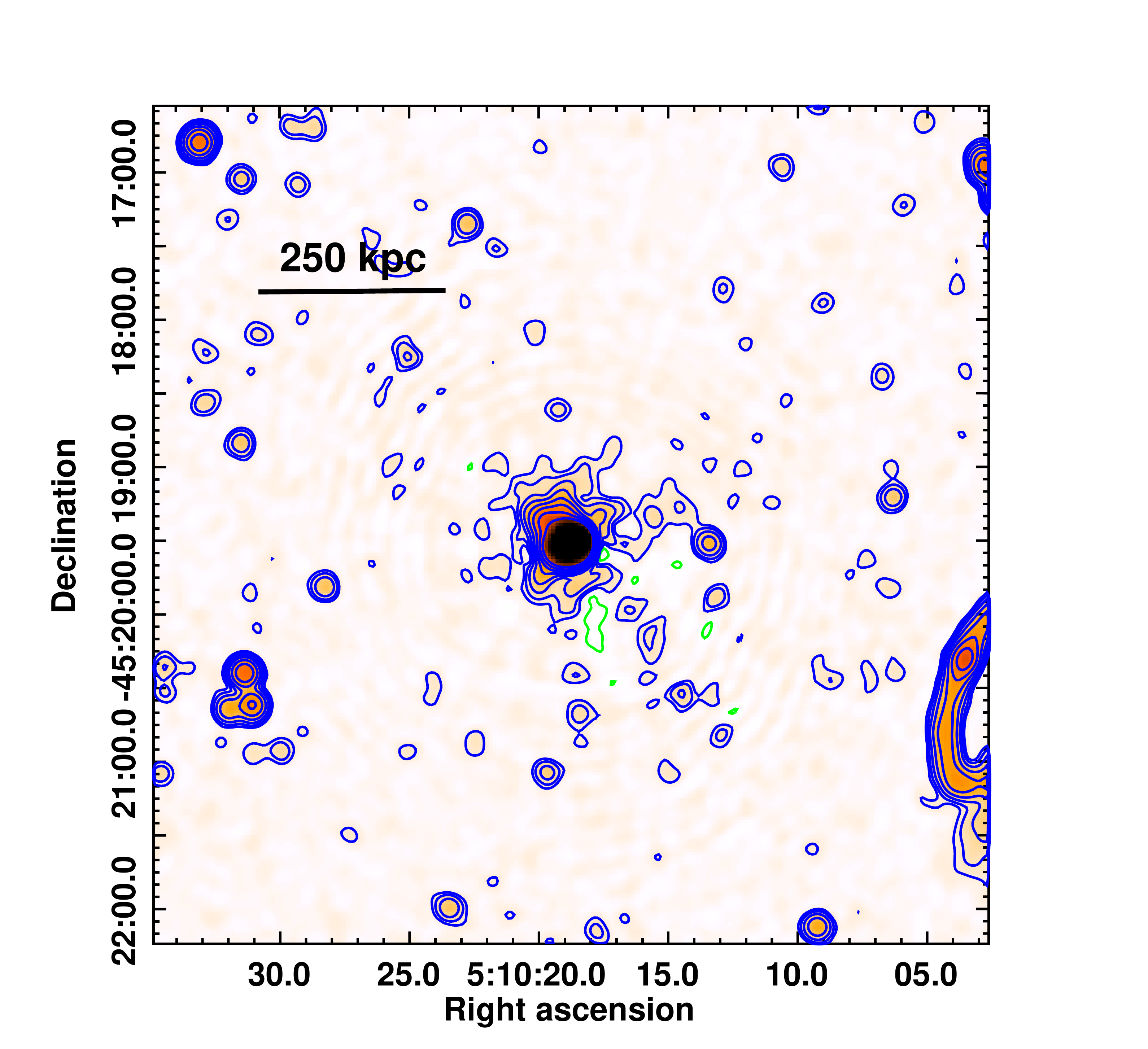}\\
        \includegraphics[height = 6.5cm]{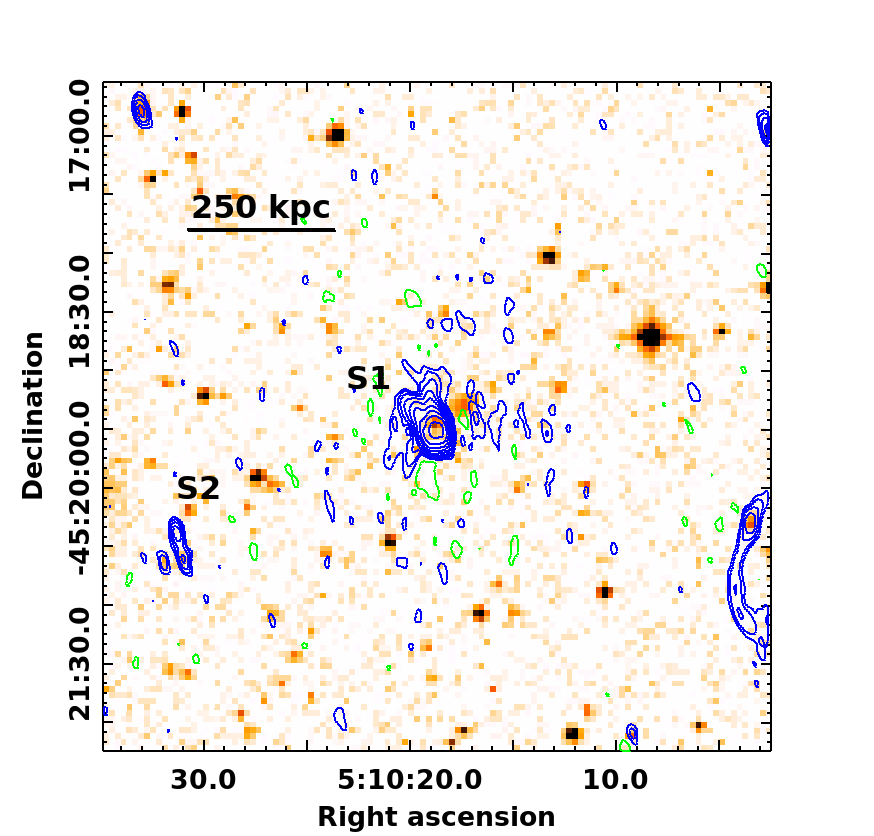}
       \includegraphics[height = 6.5cm]{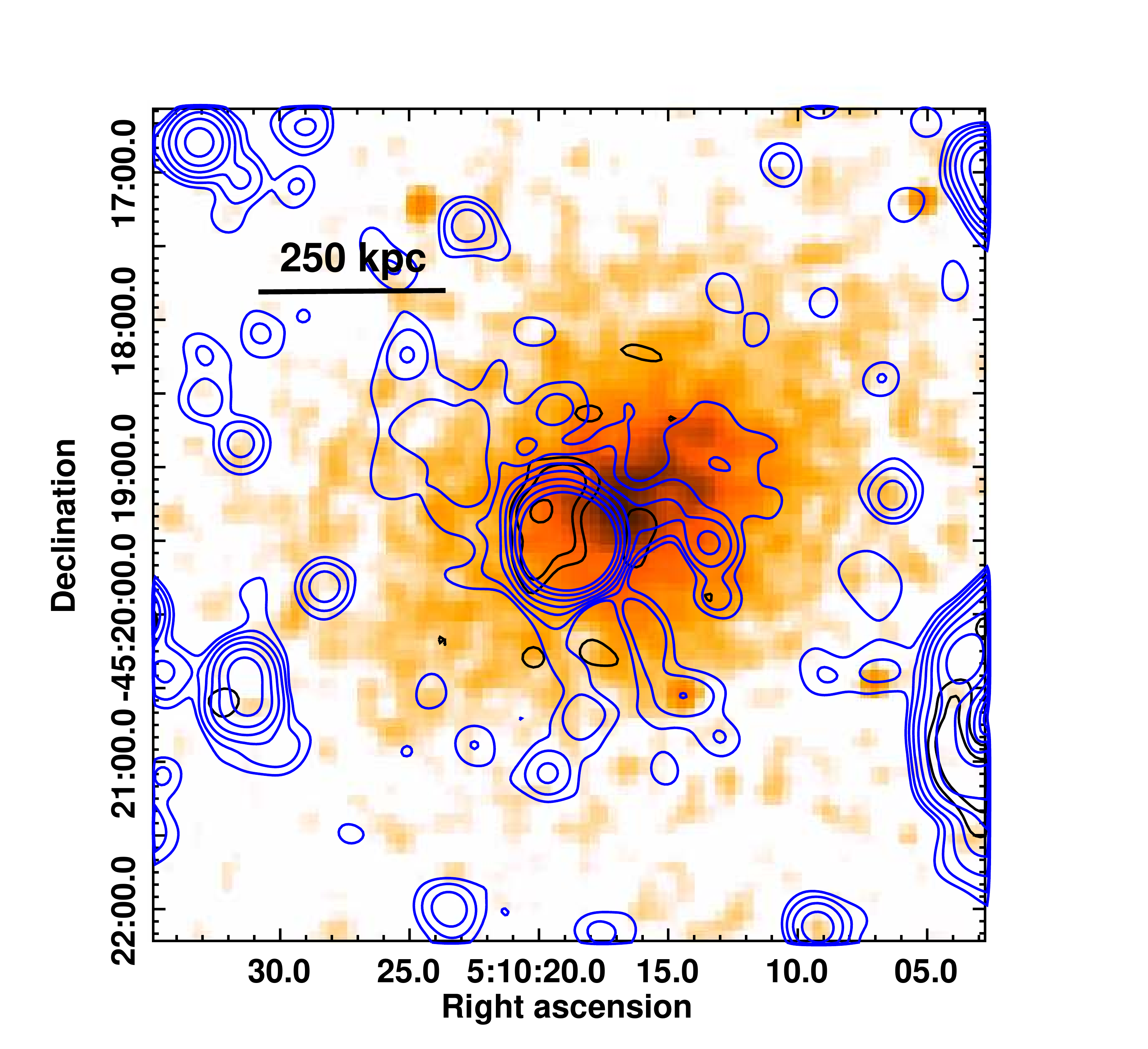}\\
        \includegraphics[height = 6.5cm]{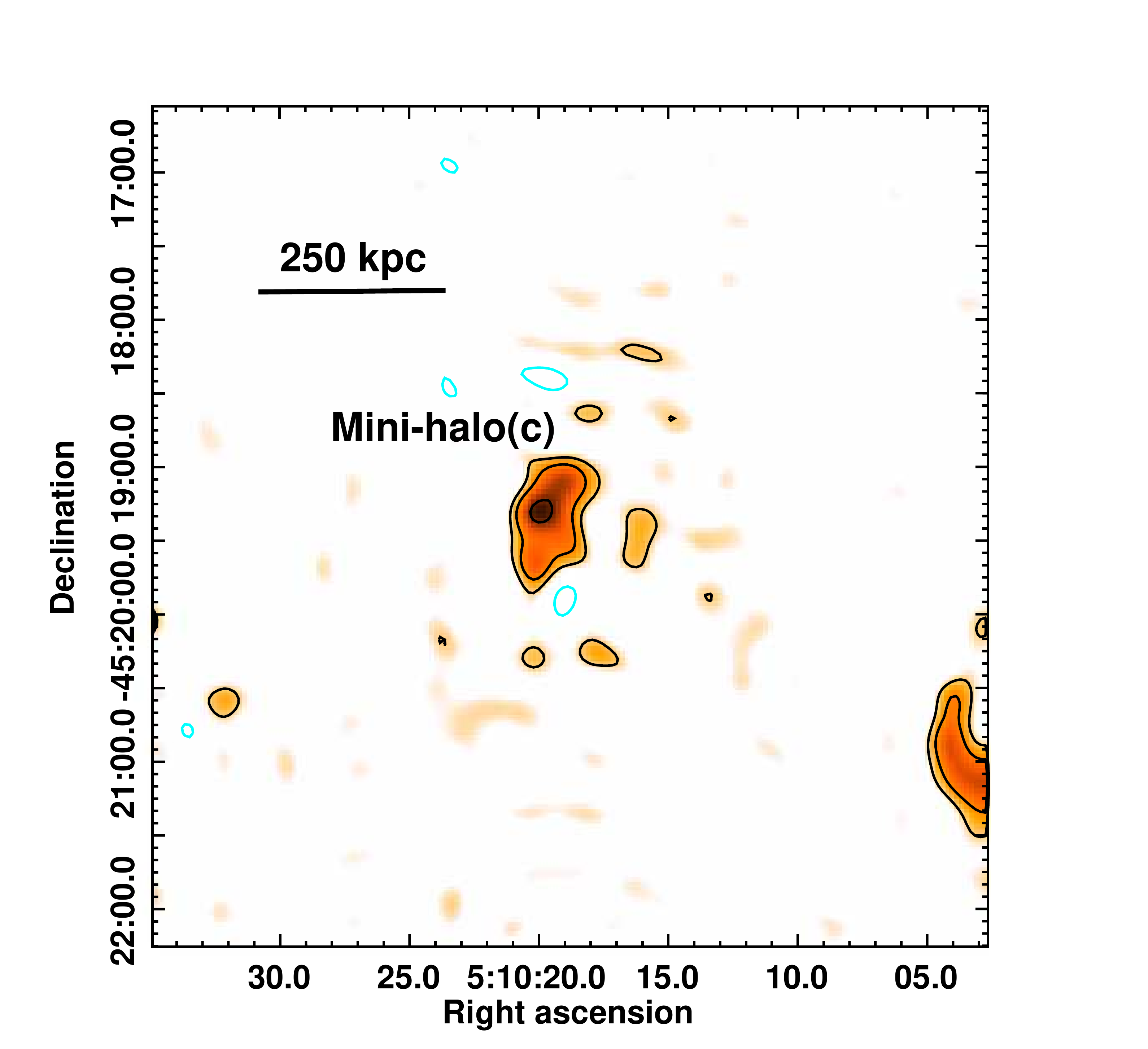}
   \includegraphics[height = 6.5cm]{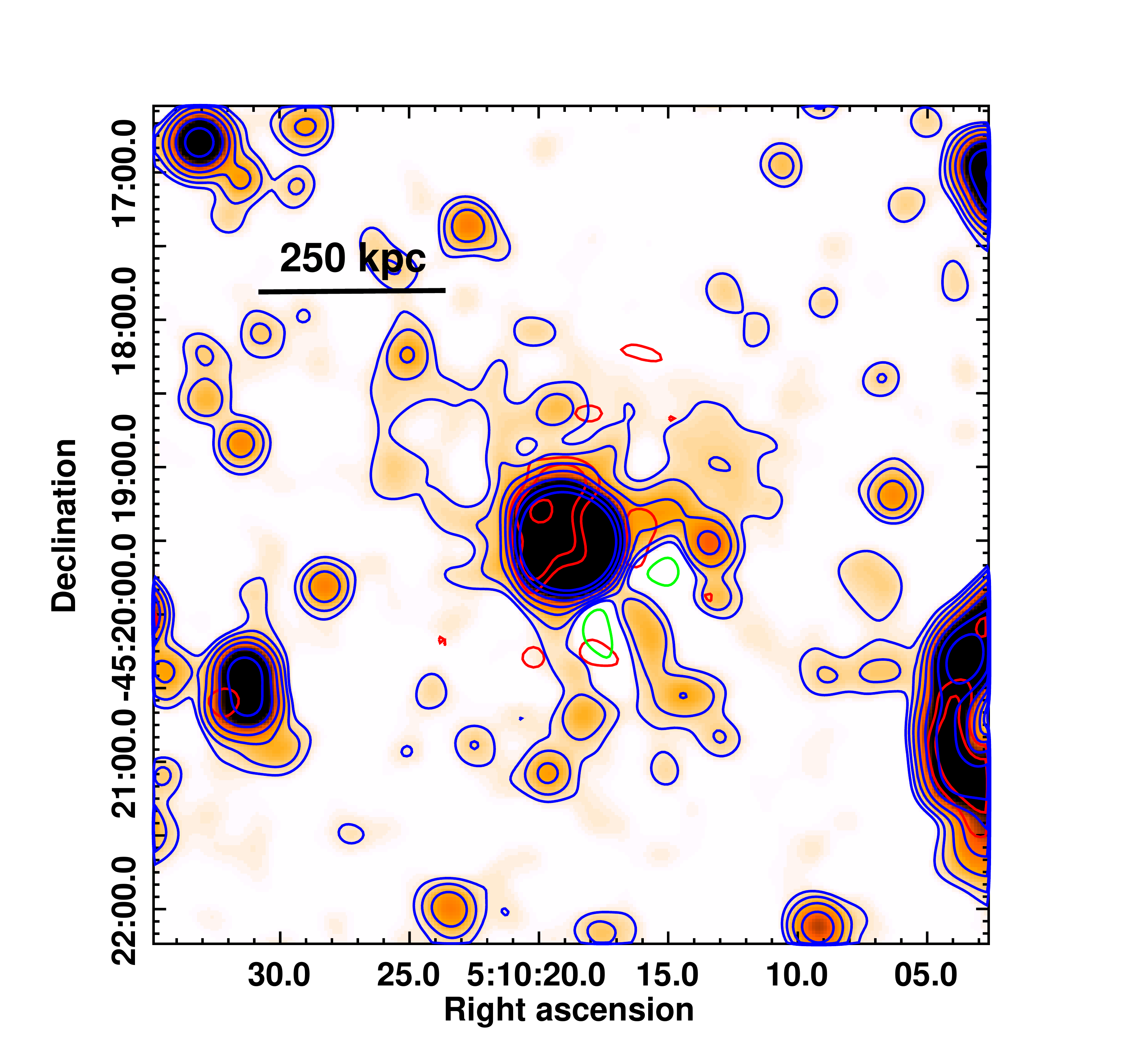}
    \caption{\textbf{A3322}: {Top left: GMRT 610 MHz image is shown in colour and contours. Contour levels are at $[-1, 1, 2, 4,...]\times3\sigma_{\mathrm{rms}}$ where $\sigma_{\mathrm{rms}} = 0.6$ mJy beam$^{-1}$.
    A circle of diameter 1 Mpc is shown. Top right: MeerKAT image of the cluster shown in colour and contours. The contour levels are at $[-1, 1, 2, 4,...]\times3\sigma_{\mathrm{rms}}$ where $\sigma_{\mathrm{rms}} = 6$ $\mu$Jy beam$^{-1}$.
    The beam size is $7.6''\times 7.5'',$ p. a. $24.2^{\circ}$. 
    Centre left: The contours from the top left panel are overlaid on the DSS2 R-band image shown in colour. Centre right: Chandra X-ray image is shown in colour overlaid with the MeerKAT $15''$ resolution image in contours shown at levels $[-1, 1, 2, 4,...]\times3\sigma_{\mathrm{rms}}$ where $\sigma_{\mathrm{rms}} = 15$ $\mu$Jy beam$^{-1}$. 
    The contours from the GMRT image after sources are subtracted are shown in black. The contours are at $[-1, 1, 2, 4,...]\times3\sigma_{\mathrm{rms}}$ where $\sigma_{\mathrm{rms}} = 0.2$ mJy beam$^{-1}$. The beam size is $17''\times13'',$ p. a. $20.4^{\circ}$. 
    Bottom left: GMRT 610 MHz image after subtracting discrete sources shown in color and contours.
    Bottom right: The positive contours from the GMRT image from the bottom left panel (red) and from the MeerKAT image from the centre right panel (blue positive and green negative) are overlaid on the MeerKAT $15''$ resolution image shown in colour. In these panels, blue, black or red show positive contours and negative contours are either in cyan or green. }
    }
    \label{G250}
\end{figure*}

\section{Results}\label{results}

We obtained GMRT images of seven out of nine clusters and used the images from the MeerKAT archive for five clusters. For the three clusters, RXC J0232.2-4420, RXC J0528.9-3927, and A3322, both GMRT and MeerKAT images were used.
The summary of the high-resolution radio images produced using the GMRT data and the archival images from the MeerKAT is given in Table~\ref{obsn}. The discrete radio sources detected within a circle of diameter 1 Mpc around the cluster center in the GMRT images are reported in Table ~\ref{gmrtradsrc}. We report on the diffuse radio emission detected in clusters RXC J0528.9-3927, A3399, A3322, and RXC J0232.2-4420. The remaining clusters with non-detected diffuse radio emission are also described. 
The lists of discrete sources found within the regions of diffuse radio emission in MeerKAT images are reported in electronic tables \footnote{\url{https://cdsarc.u-strasbg.fr/ftp/vizier.submit//SUCCESS/}}. 
The images overlaying the region used for flux density measurement and the discrete sources are shown in App.~\ref{app:meerkat}.

\subsection{\textbf{Radio halo in RXC J0528.9-3927}}\label{sec:rxcj0528}
The cluster RXC J0528.9-3927 is a massive cluster at a redshift of 0.28. It is known to have a peaked central X-ray emission \citep{2005A&A...442..827F} and a mixed (intermediate to relaxed and merging) dynamical state \citep{2017ApJ...846...51L}. A cold front located to the west of the center has been reported in this cluster \citep{2018MNRAS.476.5591B}. Recently, using pilot observations with MeerKAT, a radio halo has been reported at 1.28 GHz \citep{2021MNRAS.504.1749K}. They report the largest linear size of the halo to be {0.54 Mpc} and the flux density at 1.28 GHz to be $3.58\pm 0.26$ mJy. The radio halo is reported, which extended more to the southwest of the BCG.

Our GMRT observations at 610 MHz show the presence of an extended radio source around the central galaxy (Fig.~\ref{G244}). We made separate images of the compact and diffuse sources using cut-offs in the uv-ranges of the visibilities. An image of point sources using a uv-range $>4 k\lambda$ was made. The corresponding model was subtracted from the calibrated visibilities. We then created the image with the residual visibility file using the uv-range $< 9 k\lambda$. We confirm the presence of diffuse emission in this cluster (Fig.~\ref{G244}). The largest linear size of the diffuse emission is $300$ kpc, and it has a flux density of $9.6\pm1.0$ mJy at 610 MHz. The properties of the diffuse emission are summarised in Table ~\ref{tab:meerkatextsrc}.

The discrete radio sources found within the region of the diffuse radio emission 
are marked in Fig.~\ref{appfig:rxcj0528}.
We term this source as a mini-halo, given that the size is less than that of typical radio halos ($>500$ kpc). The spectral index of the diffuse emission between 1280 and 610 MHz is $1.3$. Here we note that with the GMRT we are  detecting a smaller extent than in MeerKAT. This is mainly due to the limited sensitivity ($\sigma_{rms}$) as the extent is within the angular scales that the GMRT can sample.

\subsection{A double relic and a halo in A3399 }\label{sec:a3399}
The cluster A3399 is a massive cluster at a redshift of 0.20 \citep{Bleem_2015ApJS..216...27B}. It was observed by \textit{Chandra} X-ray telescope as a \textit{Chandra-Planck} Legacy Program for massive clusters of galaxies (PI: Christine Jones\footnote{\url{http://hea-www.cfa.harvard.edu/CHANDRA\_PLANCK\_CLUSTERS/}}). 
Our GMRT data could not be used to produce images due to calibration failure. 

We used MeerKAT L-band observations from the MGCLS \citep{2022A&A...657A..56K}, which reveals that it hosts a double relic and a radio halo. We present MeerKAT L-band images in two different resolutions, one is native 7$''$ resolution, and another one is convolved with $15\arcsec \times 15 \arcsec$ in Fig. \ref{fig:J0637}. The high sensitivity of the MeerKAT helps us recover large-scale radio relics towards the southeast from the center of the cluster (SE relic). The largest linear size of this relic is $\sim$938 kpc within the 3$\sigma$ contour. 
There is diffuse radio emission over the central region as well. The largest linear size of the diffuse emission is 647 kpc. 
There is a diffuse radio structure towards the northwest direction that we name as the candidate NW relic (NW-Rc). It has the largest linear size of 404 kpc. The flux densities after subtraction of the discrete sources are summarised in Table ~\ref{tab:meerkatextsrc}. For the NW-R, the flux density could not be measured due to the presence of several discrete sources. 
The discrete radio sources within the region of the diffuse emission are marked in Fig.~\ref{appfig:a3399}.

\begin{figure*}
\centering
\includegraphics[trim = 2cm 1.5cm 3cm 3cm,clip,height=9cm]{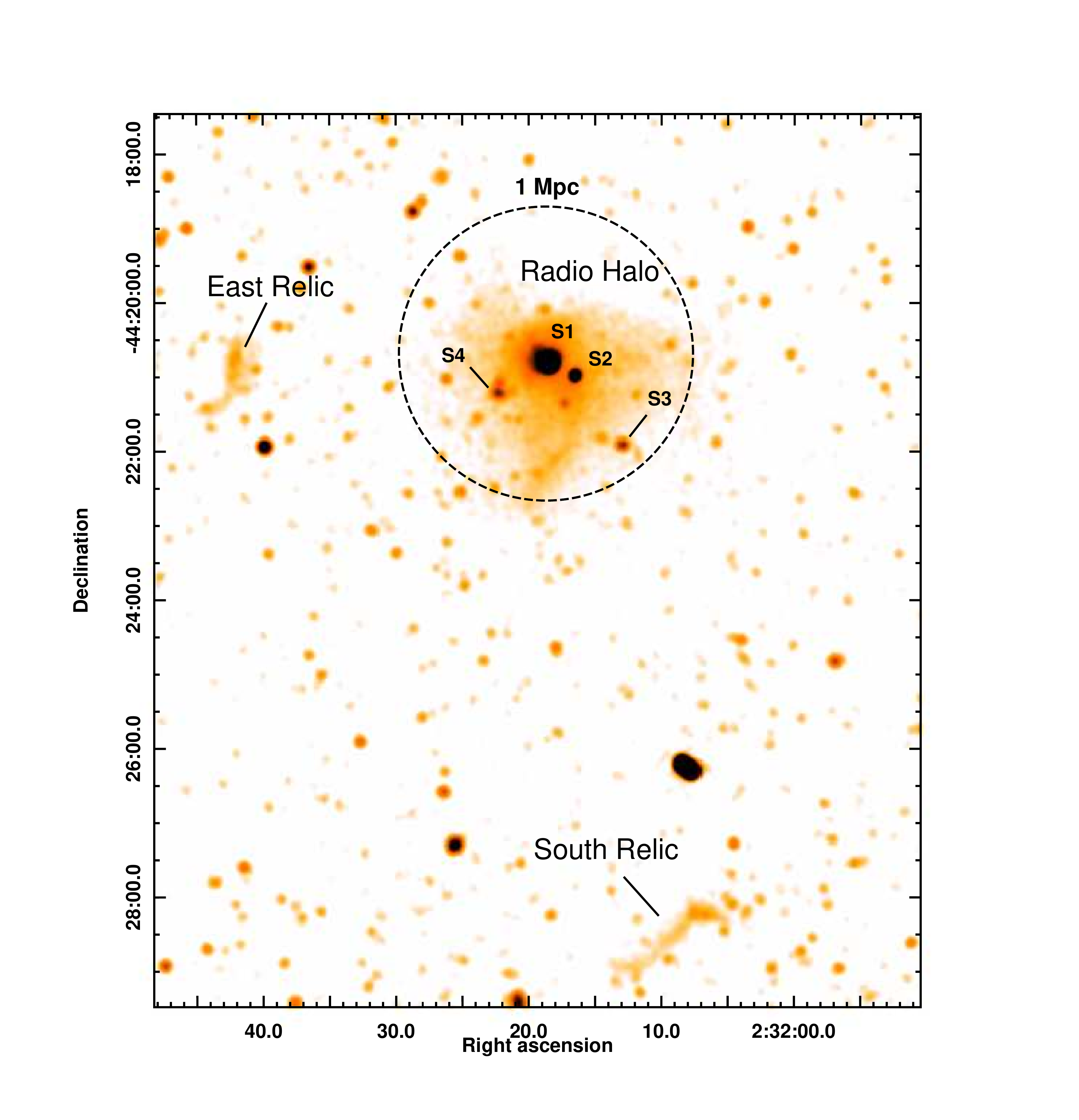}
\includegraphics[trim = 2cm 1.5cm 3cm 3cm,clip,height = 9cm]{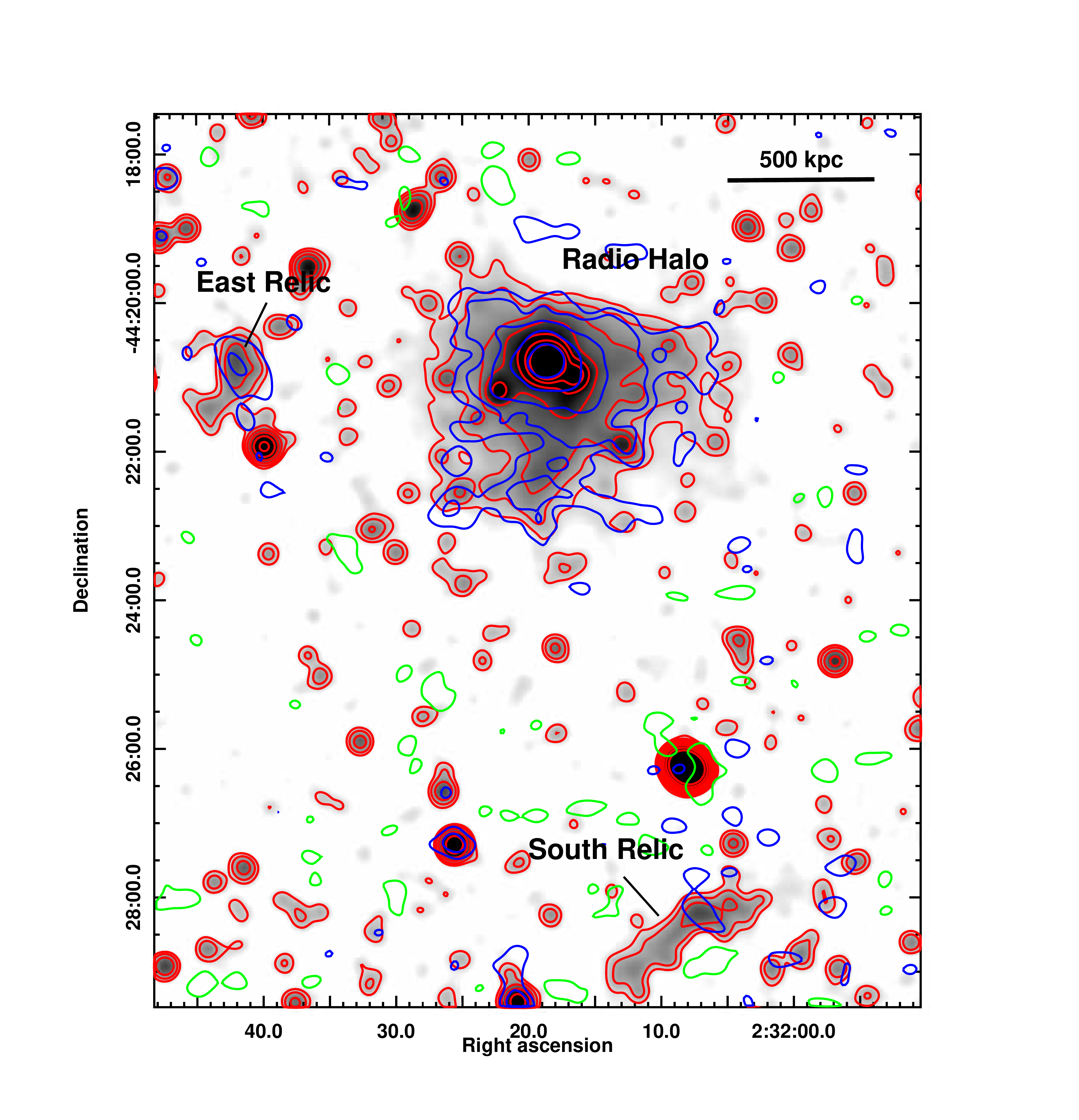}
\caption{\textbf{RXC J0232.2-4420} Left: MeerKAT high-resolution image is shown in colour scale. The source labels S1, S2, S3, and S4 from \citet{2019MNRAS.486L..80K} are shown along with that of the radio halo and the east and south relics. A circle of diameter 1 Mpc around the cluster centre is shown. Right: MeerKAT low resolution ($15''\times15''$) image is shown in colour and in contours (red). The red contour levels are  $[-1, 1, 2, 4,...]\times3\sigma_{\mathrm{rms}}$ where $\sigma_{\mathrm{rms}} = 10$ $\mu$Jy beam$^{-1}$. 
There are no negative contours at the level of $-30 \mu$Jy beam$^{-1}$. The GMRT 610 MHz contours of the radio halo from the point source subtracted image with a resolution of $20''\times20''$ from \citet{2019MNRAS.486L..80K} are shown in blue (positive) and green (negative). 
The contour levels are $[-1, 1, 2, 4,...]\times3\sigma_{\mathrm{rms}}$ where $\sigma_{\mathrm{rms}} = 0.83$ mJy beam$^{-1}$.}
\label{fig:J0232}
\end{figure*}

\begin{figure*}
\centering
\includegraphics[trim = 0cm 0cm 0cm 2cm,clip,height=7cm]{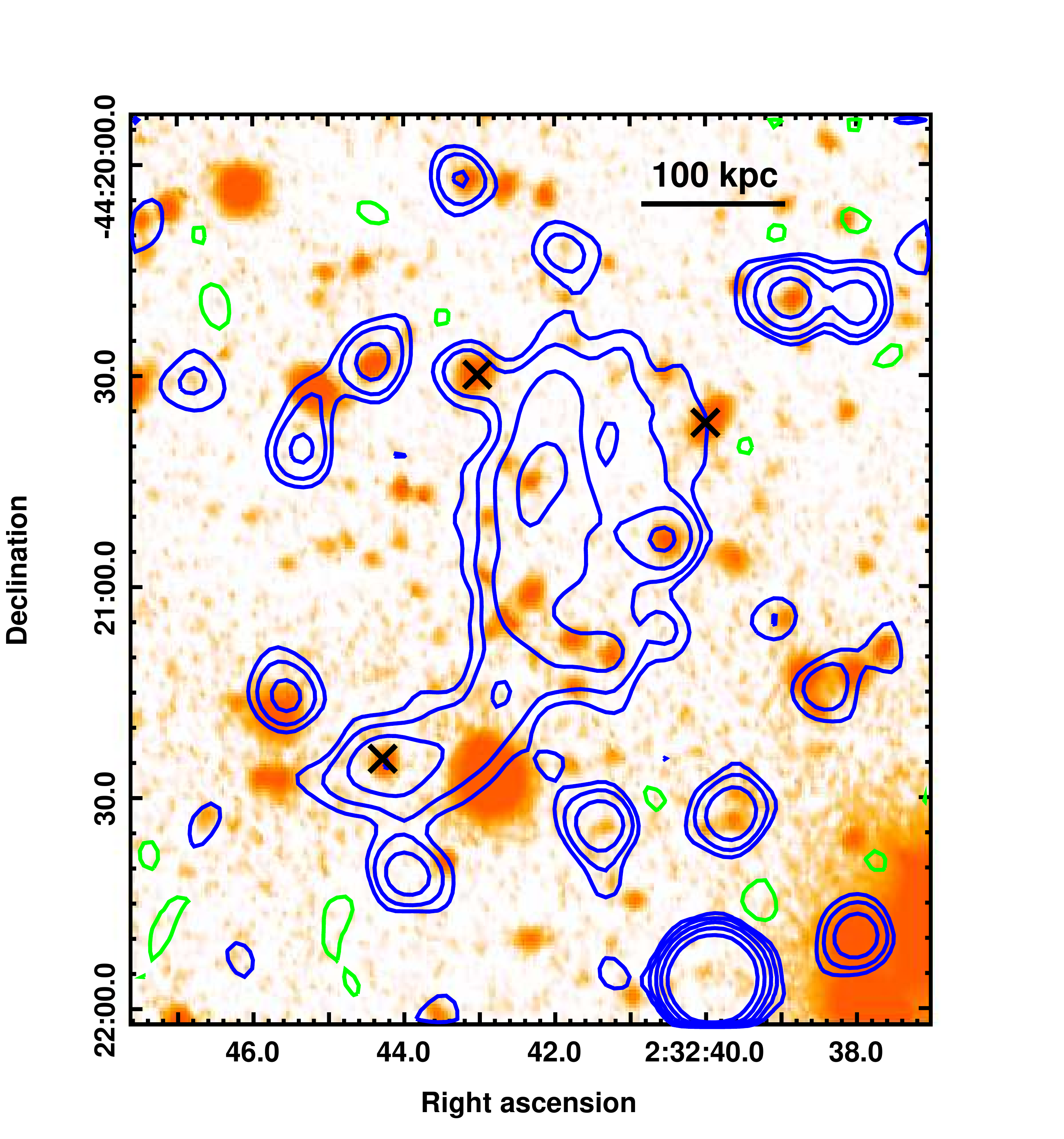}
\includegraphics[trim = 0cm 0cm 0cm 2cm,clip,height = 7cm]{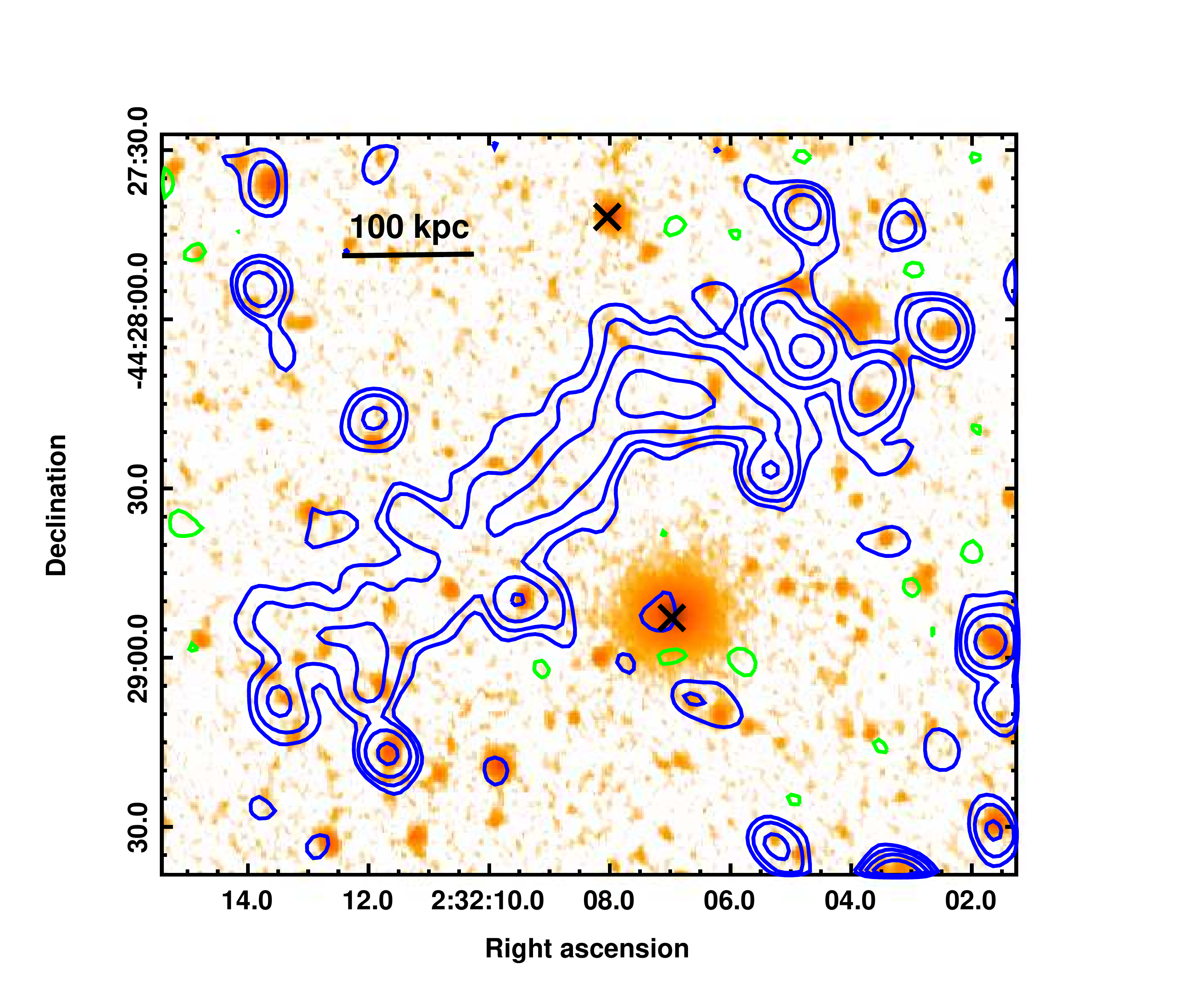}
\caption{\textbf{RXC J0232.2-4420}: Dark Energy Survey DR1 r-band image \citep{2018ApJS..239...18A} is shown in colour 
with MeerKAT high-resolution image contours overlaid. The regions of the East relic (left) and 
of the South relic (right) are shown. The MeerKAT contours are at $[-1, 1, 2, 4,...]\times3\sigma_{\mathrm{rms}}$ where $\sigma_{\mathrm{rms}} = 2.6$ $\mu$Jy beam$^{-1}$. The blue contours are positive, and green are negative. The crosses mark the positions of 
objects identified as galaxies in the NASA Extra-galactic database. 
}
\label{fig:J0232-relics}
\end{figure*}

\subsection{A mini-halo candidate in A3322}\label{sec:a3322}
The cluster A3322 is a massive cluster at a redshift of 0.20. In the GMRT 610 MHz images (Fig.~\ref{G250}, left panels) of the cluster, we find the presence of an extended component partially cospatial with the discrete source S1. After subtraction of the emission from the discrete source, the residual emission has the largest angular size of $75''$, which at the redshift of the cluster corresponds to 248 kpc. The brightest part of this source (RA$_{\mathrm{J}2000}$ 05h10m19.535s, DEC$_{\mathrm{J}2000}$ $-45^{\circ}19^{\prime}17.36^{\prime\prime}$) is offset from the brightest part of S1 (RA$_{\mathrm{J}2000}$ 05h10m18.764s, DEC$_{\mathrm{J}2000}$ $-45^{\circ}19^{\prime}30.80^{\prime\prime}$) which makes it probable that this source is separated from S1. In the MGCLS, this cluster has been considered a candidate mini-halo of size 0.26 Mpc, comparable to the size we report. 

We also present the MeerKAT 1280 MHz images of this cluster (Fig.~\ref{G250}, right panels). The overlay of the radio image on the Chandra X-ray image shows that the central source is slightly offset from the peak of the X-ray emission. The MeerKAT images show evidence of extended emission. Due to the negative regions next to the central source, the flux density of the diffuse emission could not be measured. 
The discrete radio sources near the centre are marked in Fig.~\ref{appfig:a3322}.
We consider the diffuse source to be a mini-halo candidate. The diffuse source has a flux density of $8.4\pm1.3$ mJy at 610 MHz and a 1.4 GHz radio power of $7.5\times10^{23}$ W Hz$^{-1}$ (Table~\ref{tab:meerkatextsrc}).

\subsection{Radio halo and candidate relics in RXC J0232.2-4420}\label{sec:rxcj0232}
The discovery of a radio halo in RXC J0232.2-4420 using the GMRT was reported by \citet{2019MNRAS.486L..80K}. The MeerKAT image of the cluster is shown in Fig.~\ref{fig:J0232}. The radio halo discovered with the GMRT has been detected with the MeerKAT as well. In addition, we detect two elongated filamentary diffuse features - one towards the east and one to the south of the cluster. We have labeled them as East Relic and South Relic. The East Relic is at a distance of $250''$ (1.05 Mpc), and the South Relic is at a distance of $464''$ (1.95 Mpc). The largest linear sizes of the East and the South Relics are $91''$ (380 kpc) and $140''$ (600 kpc), respectively. {We examined the Dark Energy Survey Data Release 1 r-band images and looked for identifications (Fig.~\ref{fig:J0232-relics}). We do not find any obvious optical identifications with the galaxies known in the region of the two diffuse relics.} Given their large projected distances from the cluster center, it is possible that these are sources unrelated to the cluster. The origin of such filamentary radio sources can be radio galaxies whose central activity has stopped. These are called remnant radio galaxies and are likely to be short-lived unless revived \citep[e. g.][]{ens01,2020MNRAS.496.1706S}. 
The largest extent of the radio halo in the MeerKAT image is found to be $258''\times270''$. We note that the discrete sources have not been subtracted from the MeerKAT images.The images showing the radio halo and the relics with the discrete sources are given in Fig.~\ref{appfig:rxcj0232}. 
Further detailed analysis of the diffuse sources in this cluster is deferred to a subsequent work where we will combine the MeerKAT and uGMRT observations. 

\subsection{Non-detections: RXC J1358.9-4750, PSZ1 G313.85+19.21, A3396, A3937 and A3343}\label{sec:nondet}
In this sub-section, we present the radio images of the five clusters where no diffuse emission associated with the intra-cluster medium of the clusters was found in our observations. 

\begin{figure*}
    \centering
    \includegraphics[trim = 0cm 0cm 1cm 1cm,clip,height = 7cm]{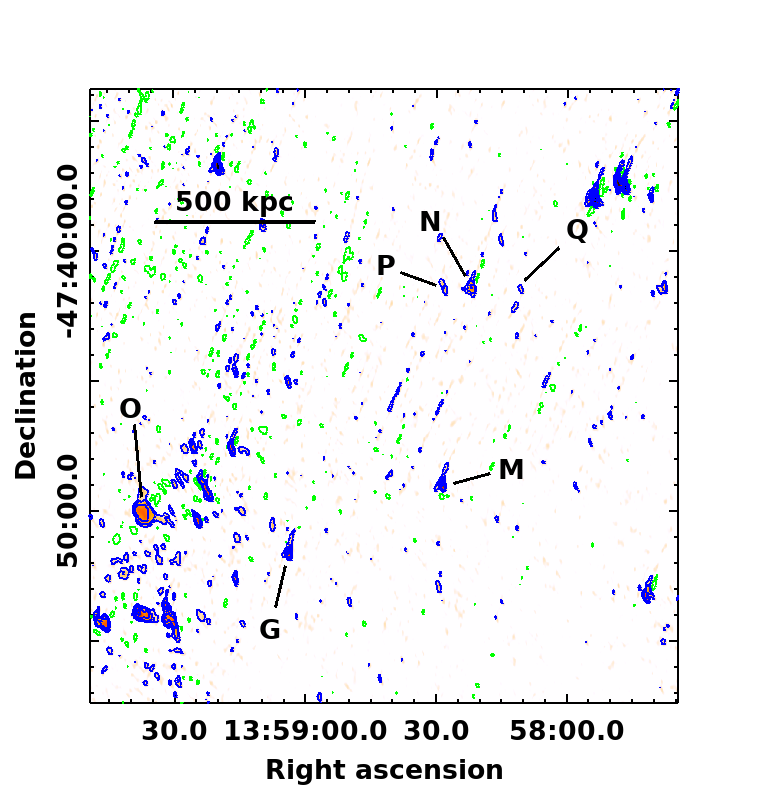}
    \includegraphics[trim = 0cm 0cm 1cm 1cm,clip,height = 7cm]{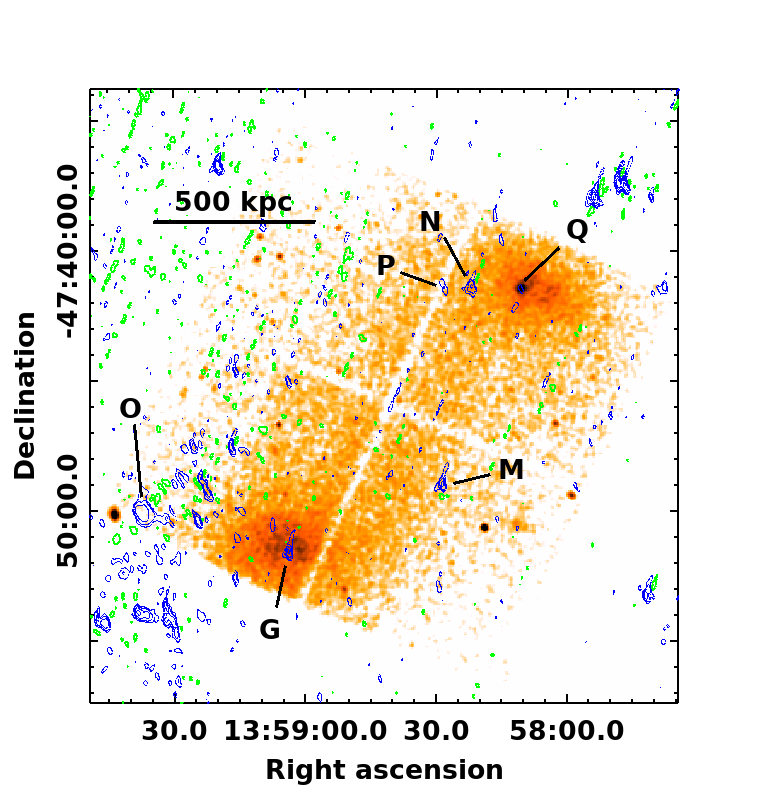}\\
   \includegraphics[trim = 0cm 0cm 1cm 1cm,clip,height = 7cm]{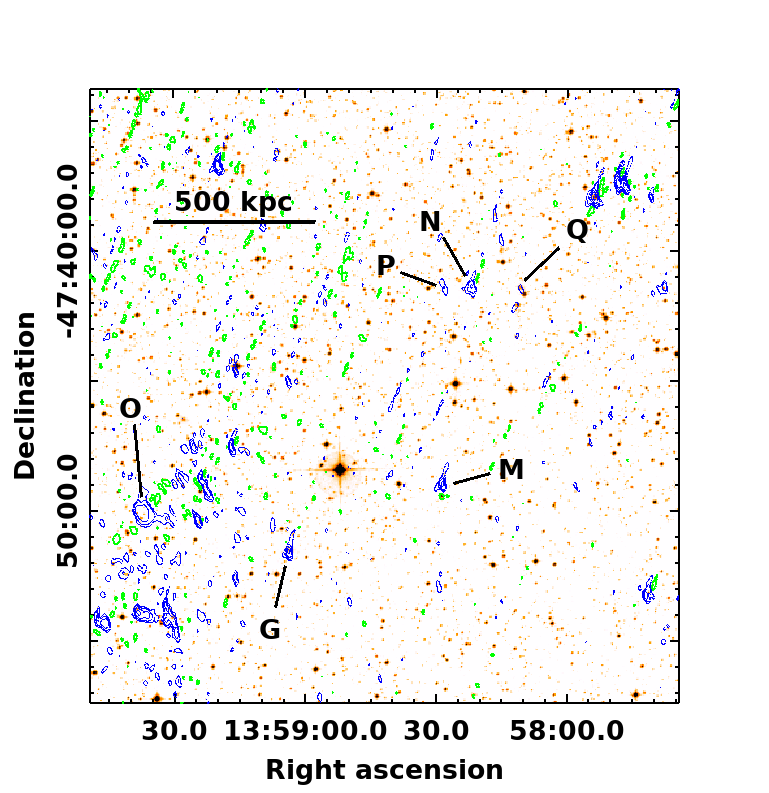}
    \caption{\textbf{RXC J1358.9-4750}: Top left: GMRT 325 MHz image shown in colour and contours. The beam size is $19.4''\times6.7''$, position angle 6.4$^{\circ}$. Contour levels are at $[-1, 1, 2, 4,...]\times3\sigma_{\mathrm{rms}}$ where $\sigma_{\mathrm{rms}} = 0.5$ mJy beam$^{-1}$. The positive contours are blue and negative in green. Top right: The contours, same as in the top left panel, are overlaid on the X-ray image of the cluster shown in colour. Bottom: The contours, same as in the top left panel, are overlaid on the DSS2 R-band image shown in colour. The discrete sources are labeled.}
    \label{appfig:rxcj1358}
\end{figure*}

RXC J1358.9-4750 has been identified to be a cluster in an early phase of the merger based on the Suzaku and XMM-Newton observations \citep{2015PASJ...67...71K}. The X-ray emission has two main clumps with temperatures $5.6\pm0.2$ keV and $4.6\pm0.2$ keV separated by 1.2 Mpc that are joined by a bridge. The bridge is a high-temperature region ($>9$ keV) and indicates shock heating in an early phase of a head-on merger. Radio observations with the Australia Telescope Compact Array at 16 cm show absence of diffuse emission and a 1.4 GHz upper limit of $1.1\times10^{22}$ W Hz$^{-1}$ has been placed \citep{2018PASJ...70...53A}. In the 325 MHz image with the GMRT (Fig.~\ref{appfig:rxcj1358}), we detect the source G, which was labeled in \citet{2018PASJ...70...53A} and in addition label the sources M, N, O, P, and Q (Table~\ref{gmrtradsrc}). The source G is at the core of the southern cluster, and we detect source Q, which is at the core of the northern cluster. The source O located to the east of the southern cluster is extended and shaped like a bubble. We did not find any identifications in optical or IR bands for M, O, and P that we list. Source N is associated with an X-ray source, and sources G and Q have identifications in the WISE survey. 

\begin{figure*}
    \centering
    \includegraphics[height = 8cm]{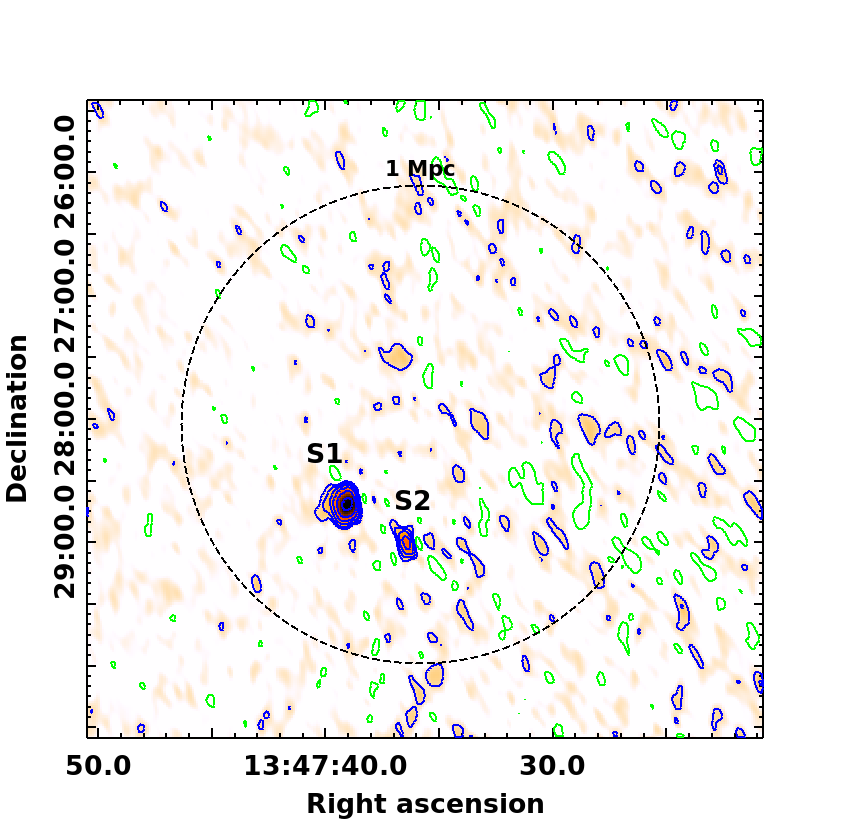}
    \includegraphics[height = 8cm]{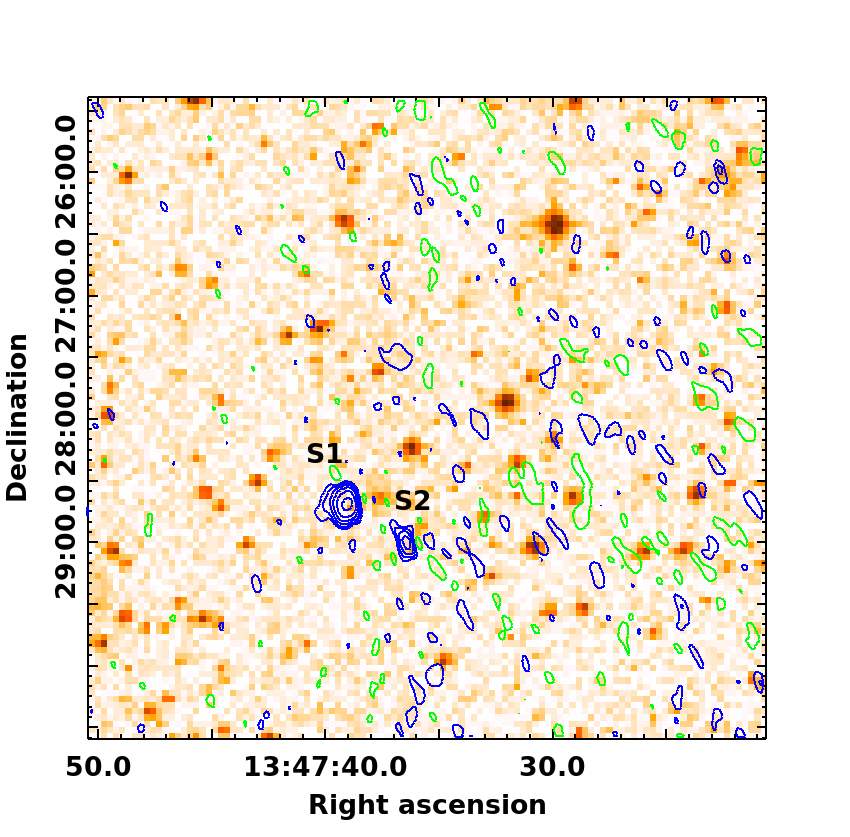}
    \caption{\textbf{PSZ2G313.84+19.21}: Left: GMRT 610 MHz shown in colour and contours. The beam size is $10.7''\times4.0''$, position angle 14.7$^{\circ}$. Contour levels  are $[-1, 1, 2, 4,...]\times3\sigma_{\mathrm{rms}}$ where $\sigma_{\mathrm{rms}} = 0.06$ mJy beam$^{-1}$. The positive contours are blue and negative in green. A circle of diameter 1 Mpc around the cluster centre is shown. Right: The contours from the left panel are overlaid on the DSS2 R-band image shown in colour. The sources S1 and S2 are labeled in both panels.}
    \label{appfig:g313}
\end{figure*}

PSZ1 G313.85+19.21 is a cluster at a redshift of 0.28 where we do not detect diffuse radio emission (Fig.~\ref{appfig:g313}). There are two discrete radio sources for which WISE identifications were found (Table ~\ref{gmrtradsrc}). 

\begin{figure*}
    \centering
    \resizebox{\textwidth}{!}{
    \begin{tabular}{c|c}
         \includegraphics[width=\columnwidth]{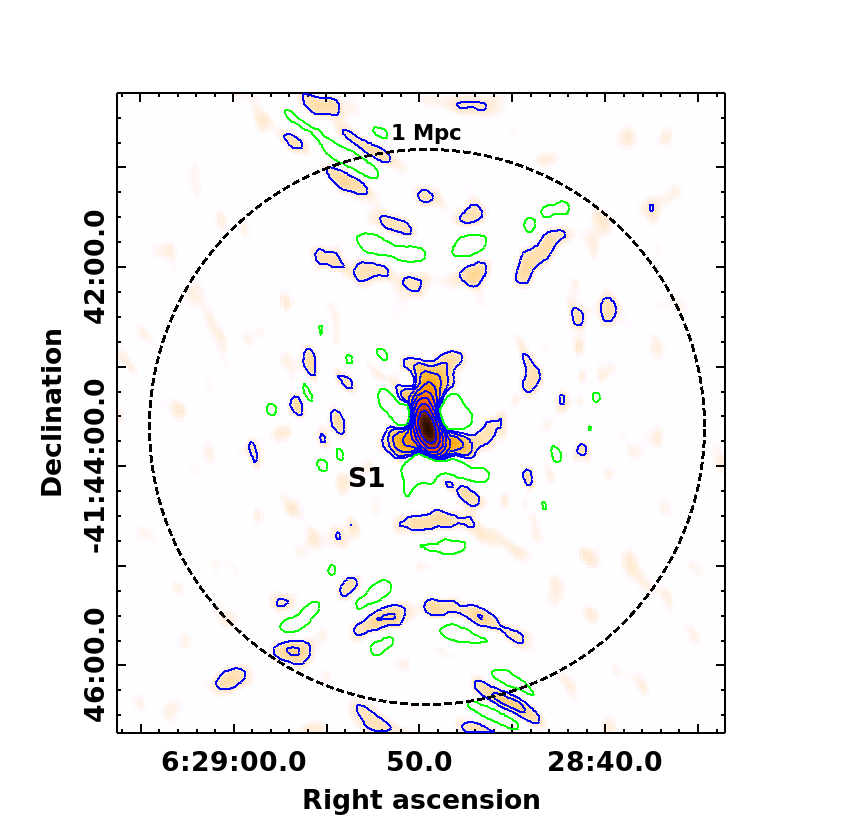} & \includegraphics[width=\columnwidth]{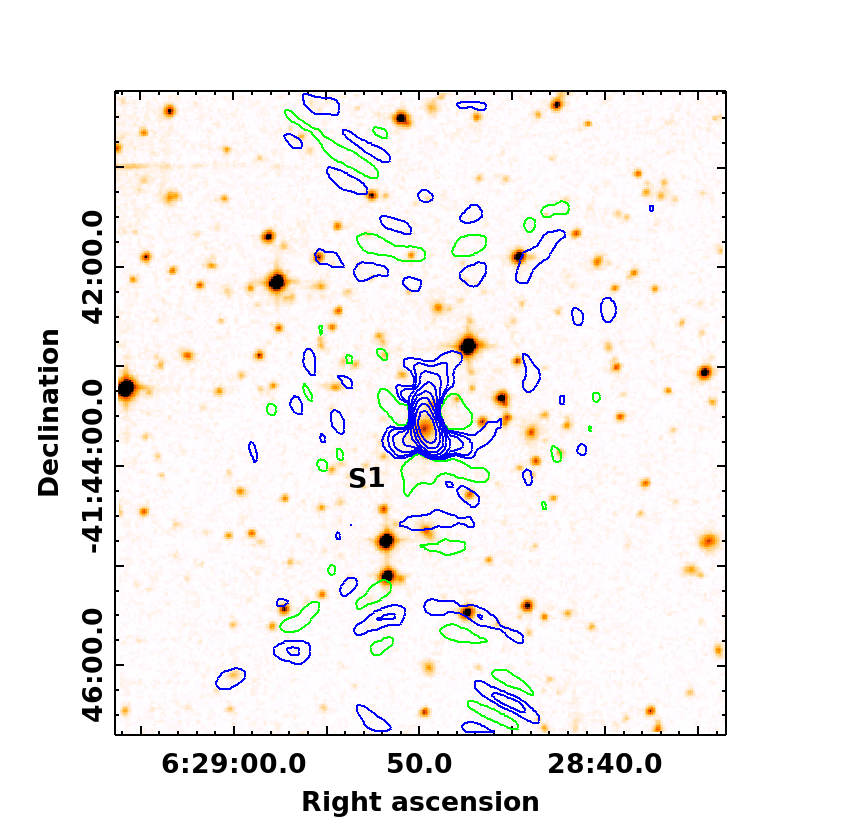} \\
    \end{tabular}
    }
    \caption{\textbf{A3396}: Left: GMRT 325 MHz image shown in colour and contours. The beam size is $14.4''\times7.5''$, position angle 6.3$^{\circ}$. Contour levels are at $[-1, 1, 2, 4,...]\times 3\sigma_{\mathrm{rms}}$ where $\sigma_{\mathrm{rms}} = 2.0$ mJy beam$^{-1}$. The positive contours are blue and negative in green. A circle of diameter 1 Mpc around the cluster centre is shown. Right: The contours from the left panel are overlaid on the DSS2 R-band image. The source S1 is labeled in both panels.}
    \label{appfig:3396}
\end{figure*}

A3396 is a cluster at a redshift of 0.176 
with a temperature of $6.312^{+0.695}_{-0.659}$ keV \citep{2020A&A...636A..15M}. Due to a bright source at the center of this cluster, the dynamic range in the surrounding region is compromised (Fig.~\ref{appfig:3396}). We do not detect other sources in the vicinity of the central bright source associated with the central galaxy.

\begin{figure*}
    \centering
    \includegraphics[trim = 2cm 0cm 2cm 3cm,clip,height = 7cm]{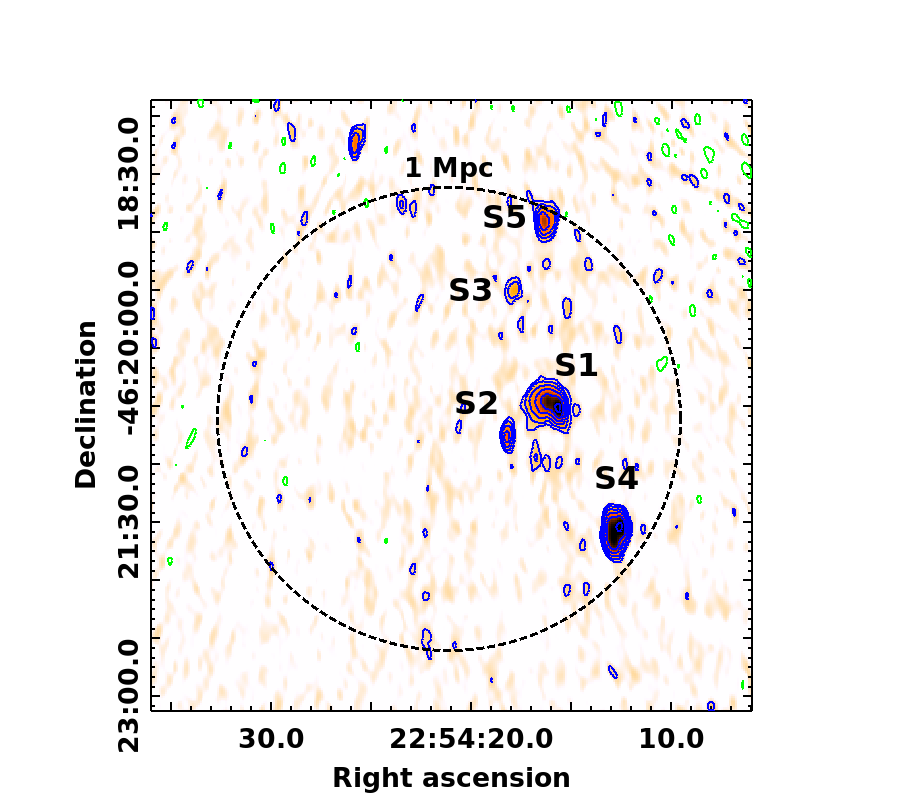}
    \includegraphics[trim = 2cm 0cm 2cm 3cm,clip,height = 7cm]{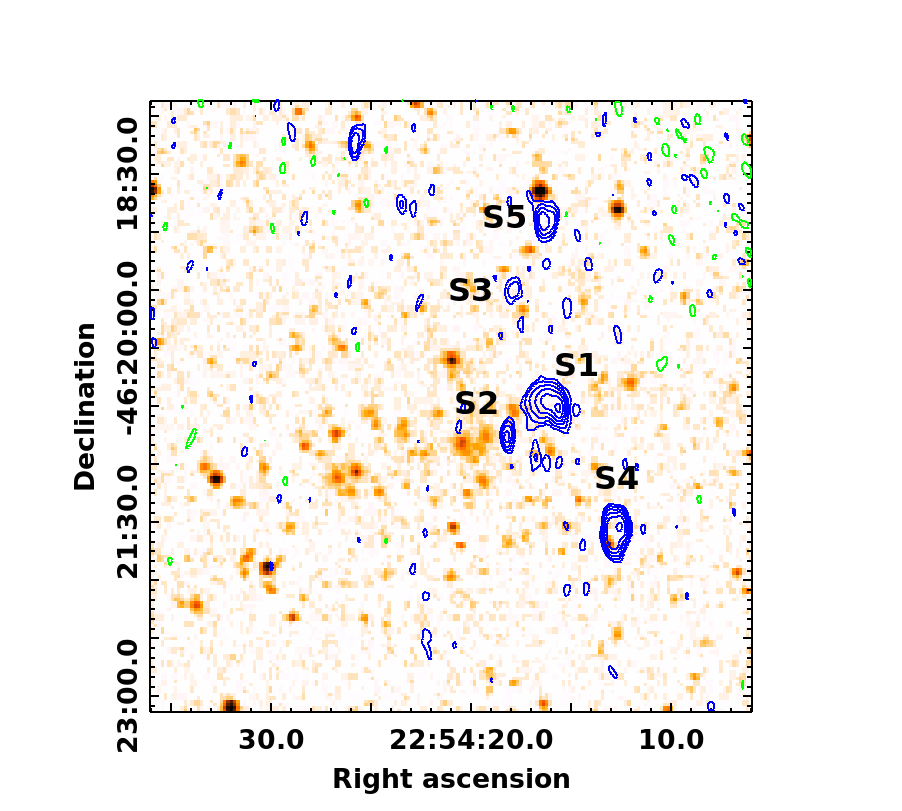}
    \caption{\textbf{A3937}: GMRT 610 image (left) and the contours of the same overlaid on DSS optical R-band image (right). Contour levels in both the panels are $[-1, 1, 2, 4,...]\times 3\sigma_{\mathrm{rms}}$ where $\sigma_{\mathrm{rms}} = 0.07$ mJy beam$^{-1}$. Positive contours are shown in blue and negative in green. The beam size is $8.6''\times3.6''$, position angle 4.9$^{\circ}$. A circle of diameter 1 Mpc around the cluster centre is shown in the left panel.}
    \label{appfig:a3937}
\end{figure*}

A3937 is a cluster at a redshift of 0.27. There are five discrete sources in the central Mpc; however, the identification of these sources in other observing bands could not be established (Fig.~\ref{appfig:a3937}, Table~\ref{gmrtradsrc}). We did not find information on the dynamical state of this cluster in the literature.

\begin{figure*}
\centering
\begin{tabular}{lccr}
\includegraphics[width=\columnwidth]{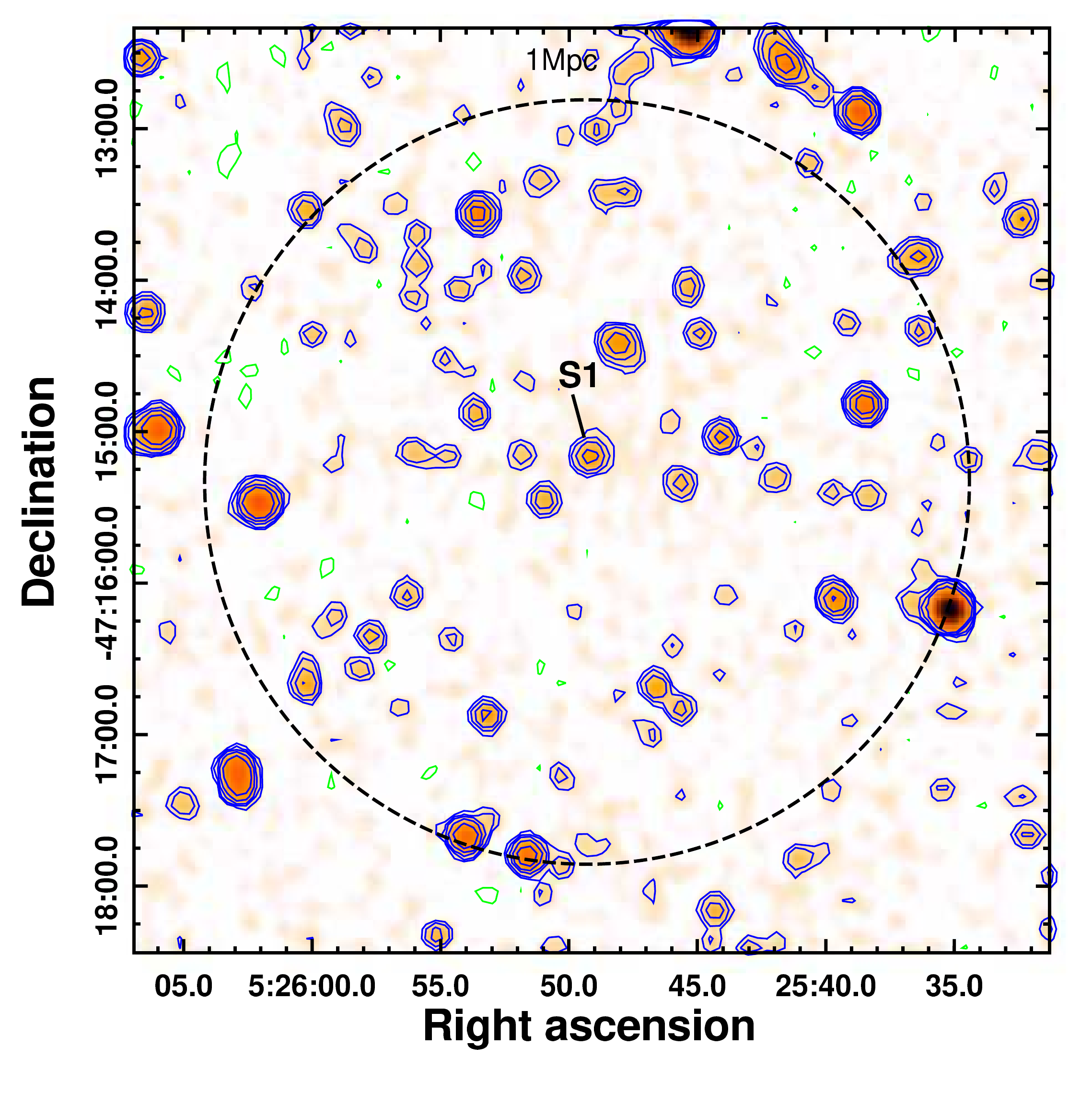} &
\includegraphics[width=\columnwidth]{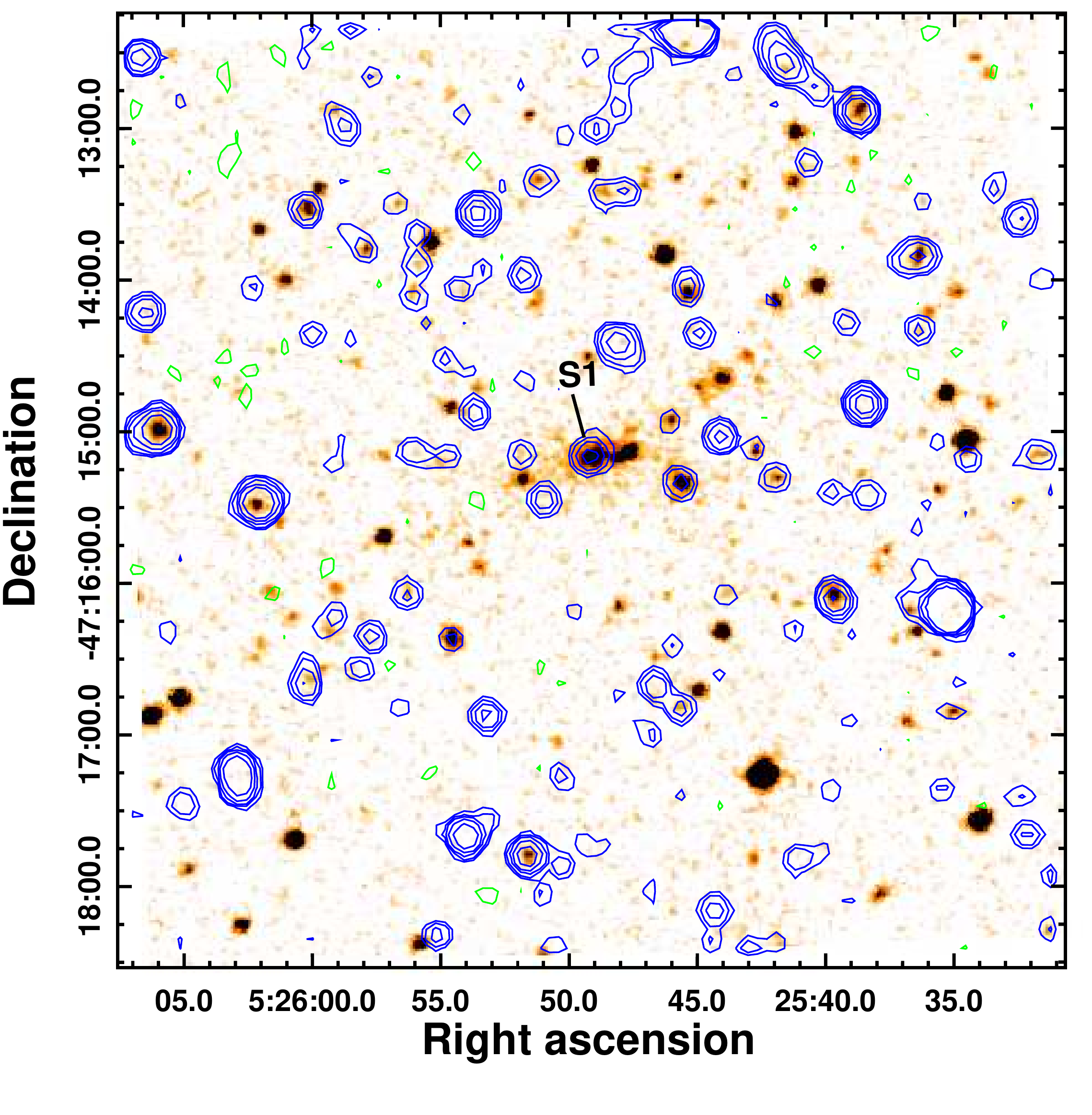} \\
\end{tabular}
\caption{\textbf{A3343}: Left: The MeerKAT 1283 MHz radio image shown in colour and contours. Contour levels are at [-1, 1, 2, 4, 8, 16, ...]$\times 3\sigma_{\mathrm{rms}}$, where $\sigma_{\mathrm{rms}} = 3 \mu$ Jy beam$^{-1}$ (beam = $8.1''\times7.7''$). Negative contours are shown in green and positive in blue. A circle of diameter 1 Mpc around the cluster centre is shown. The source S1 associated with the central galaxy is labelled. Right: The DSS2 r-band optical image overlaid with the contours of 1283 MHz  MeerKAT radio image (blue). Contour levels are same as contours in the left panel.}
\label{appfig:a3343}
\end{figure*}

A3343 is a cluster located at a redshift of 0.191. The Brightest Cluster Galaxy identified with the 2MASSX source J$05254904-4715093$ has been classified as an elliptical galaxy showing signatures of interaction \citep{2009AJ....137.4795C}. There is a radio source associated with this galaxy which is detected in the MeerKAT image and has a flux density of $0.14\pm0.01$ mJy (Fig.~\ref{appfig:a3343}). We do not find a signature of diffuse emission in this cluster. It has been classified as a relaxed cluster by \citet{2017ApJ...846...51L}.

\begin{table*}
	\centering
	\caption{Radio sources in the cluster fields are found in the GMRT images. Notes: The source G in RXC J1358.9-4750 is according to the labels A to K in \citet{takuya18} and the rest which are new are named M onwards. The labels for the discrete sources in other clusters are given as S1, S2, S3, and so on. {\bf The identification in other bands and remarks are provided in the last column.}
	}
	\label{gmrtradsrc}
	\resizebox{\textwidth}{!}{
	\begin{tabular}{cccccl} % four columns, alignment for each
		\hline
Cluster/Source	& RA$_{J2000}$	& Dec$_{J2000}$&Freq. ($\nu$)	& $S_{\nu}$	& Identification/ \\
Label        &  hh mm ss     & $^{\circ}$ $'$ $''$      &MHz&   mJy                 &Remark\\
\hline
RXC J1358.9-4750& & & &\\
G&13 59 03.746 &-47 51 35.63 &325&$22.7\pm2.9$ &WISEA J135903.81-475131.1\\
M&13 58 28.548 &-47 48 49.92 &325&$21.9\pm2.8$ &-\\
N& 13 58 21.985&-47 41 24.63 &325&$20.4\pm3.5$ &1RXS J135821.7-474126\\
O&13 59 37.104 &-47 51 35.63 &325&$106\pm15$ &-\\
P&13 58 28.278 &-47 41 24.12 &325&$9.1\pm1.7$ &-\\
Q&13 58 10.676 &-47 41 27.84 &325&$9.9\pm2.1$ &WISEA J135810.81-474124.7\\
		\hline
PSZ2 G313.84+19.21& & && &\\
	S1&13 47 39.070 & -42 28 41.75 &610&$14.2\pm0.5$ &WISEA J134738.77-422843.4\\
S2&13 47 36.434 &-42 29 00.42 &610& $2.4\pm0.2$ &WISEA J134736.24-422900.6\\
		\hline
RXC J0528.9-3927& & && &\\
S1&05 28 53.061 &-39 28 15.21 &610&$10.6\pm1.1$ &RBS 0653 (QSO)\\
S2&05 28 57.795 &-39 27 53.81 &610&$1.3\pm0.2$ &WISEA J052857.81-392755.1\\
S3&05 28 35.159 &-39 27 53.81  &610&$0.4\pm0.1$ &WISEA J052844.93-392731.4\\
S4&05 28 56.155 &-39 27 20.37  &610&$0.3\pm0.1$ &SSTSL2 J052856.08-392721.0\\
		\hline
		A3396&& & & &\\
S1&06 28 49.578 &-41 43 36.64 &325&$2574\pm370$ &WISEA J062849.75-414337.2\\
		\hline
A3322& & & &&\\
S1&05 10 18.764 &-45 19 30.80 & 610&$186\pm19$&WISEA J051018.77-451927.5\\
S2&05 10 31.162 &-45 20 28.43 & 610&$3.9\pm0.6$&WISEA J051031.13-452036.9\\
		\hline
A3937& & & &&\\
S1& 22 54 15.919 &-46 20 29.24 &610& $24.6\pm2.9$&-\\
S2&22 54 18,197 &-46 20 45.48 &610&$2.8\pm0.4$ &-\\
S3&22 54 17.884 &-46 19 29.88 &610&$11.3\pm0.1$ &-\\
S4&22 54 12.769 &-46 21 34.67 &610& $30.6\pm3.4$ &-\\
S5& 22 54 16.323 &-46 18 54.21 &610&$6.6\pm0.8$ &-\\
		\hline
	\end{tabular}
	}
\end{table*}

\begin{table*}
    \centering
    \caption{The flux densities of the diffuse sources were detected using GMRT and MeerKAT observations. The columns are (1) Cluster name, (2) emission type (RH$=$ Radio Halo, MH$=$ mini-halo, R$=$ Relic, SE$=$ Southeast, NW$=$ Northwest, E$=$ East, S$=$ South, c$=$ candidate), (3) flux density at 1283 MHz, (4) Largest linear size at 1283 MHz, (5) flux density at 610 MHz, (6) LLS at 610 MHz and (7) power at 1.4 GHz (extrapolated and k-corrected assuming $\alpha=1.2$, except for RXC J0528.9-3927 where $\alpha=1.6\pm0.4$ is used). Notes: For A3322, the flux density of the MHc could not be measured (see Sec.~\ref{sec:a3322}).}
    \resizebox{\textwidth}{!}{
    \begin{tabular}{ccccccccc}
\hline
(1) & (2)& (3) &(4) & (5)& (6)&(7) \\
 Cluster	& Type & $S_{1283 \mathrm{MHz}}$ & LLS$_{1283\mathrm{MHz}}$ &  $S_{610 \mathrm{MHz}}$& LLS$_{610\mathrm{MHz}}$ & $P_{1.4\ \mathrm{GHz}}$\\
  &   &   (mJy) & (kpc)& (mJy) & (kpc) &($10^{23}$ W Hz$^{-1}$) \\
\hline
 RXC J0528.9-3927 & MH & $3.0 \pm 0.9 $ & 710 &$9.6\pm1.0$&300& $7.4\pm2.1$\\
 \hline
  A3399  & SE-R & $ 9.5 \pm 1.1$ & 940 & -&-&$10.2 \pm 1.2$\\
    & RH &  $ 2.7 \pm 0.5 $ & 650&-& -& $2.8 \pm 0.5$ \\
    & NW-Rc & - & 400&-&- & -\\
    \hline
     A3322 & MHc   & - &-& $8.4\pm1.3$  & 250 & $7.0\pm0.2$ \\
  \hline
 RXC J0232.2-4420  & RH &  $ 9.8 \pm 2.8 $ & 1100 &$52\pm5$&$800$& $23 \pm 6$\\
   & E-R & $ 0.5 \pm 0.1$ & 380&-&- & $1.2 \pm 0.2$ \\
   & S-R &  $0.5 \pm 0.2$ & 600 &-&-& $1.2 \pm 0.4 $\\

  \hline
    \end{tabular}
    }
    \label{tab:meerkatextsrc}
\end{table*}

\section{Discussion} \label{discussion}

\subsection{Occurrence fraction of diffuse radio emission}
The recent survey of 75 massive galaxy clusters ($>6\times10^{14} \mathrm{M}_{\odot}$) by \citet{cuciti21b} shows that the fraction of radio halos in such clusters is expected to be 
$0.37\pm0.02$ and even higher, $0.67\pm0.06$ in the sub-sample with masses $>8\times10^{14} \mathrm{M}_{\odot}$. The SUCCESS sample of clusters chosen to have masses $>5\times10^{14}$ M$_{\odot}$ thus holds a high chance for detection of radio halos. In the sub-sample presented here, we have detected diffuse sources in 4 out of 9 clusters and in particular, radio halos in two clusters. Thus the fraction of radio halos is $22\%$. The fraction is $22\%$ for mini-halos, and $11\%$ for radio relics. We do note that the sample here is small. The non-detections are likely due to the limitation of sensitivity and dynamic range of the observations in the case of A3396 and PSZ2G313.84+1921. In the case of A3343 and A3396, the masses of the clusters are less than $5\times10^{14}$ M$_{\odot}$, where the likelihood of detecting radio halos drops 
\citep{cuciti21b}.
With more sensitive observations with the Upgraded GMRT and MeerKAT, we may find more of the southern clusters to be detections given the trend observed with the sensitive observations with the Low-Frequency Array (LOFAR) at 144 MHz \citep{weeren21,gennaro20}. The classification of mini-halos may also change if the sources turn out to be more extended, as has been the case, for example, with \citet{2017A&A...603A.125V}.

With the revision of masses of the clusters with the more recent Planck catalogue \citep{2016A&A...594A..27P}, we find that two of the clusters in our sub-sample of nine (A3396 and A3343, Table ~\ref{clusinfo}) turned out to have masses lower than the selection criterion ($>5\times10^{14}$ M$_{\odot}$). 
This might also lead to lower halo fractions as compared to other samples of 
clusters with masses $>6\times10^{14} \mathrm{M}_{\odot}$.

\subsection{Radio power Vs Mass}
\begin{figure*}
    \centering
         \includegraphics[height = 7cm]{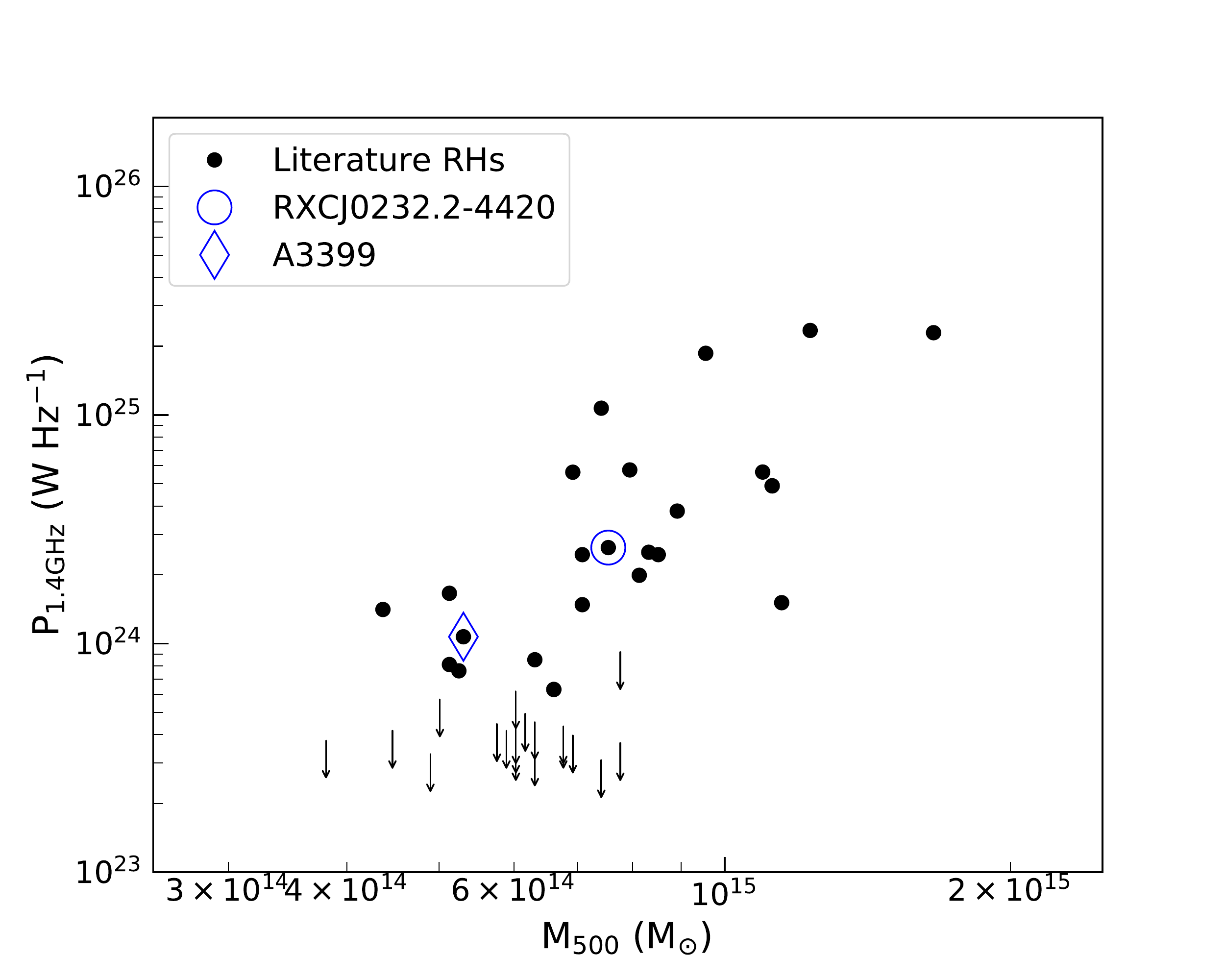}
                  \includegraphics[height = 7cm]{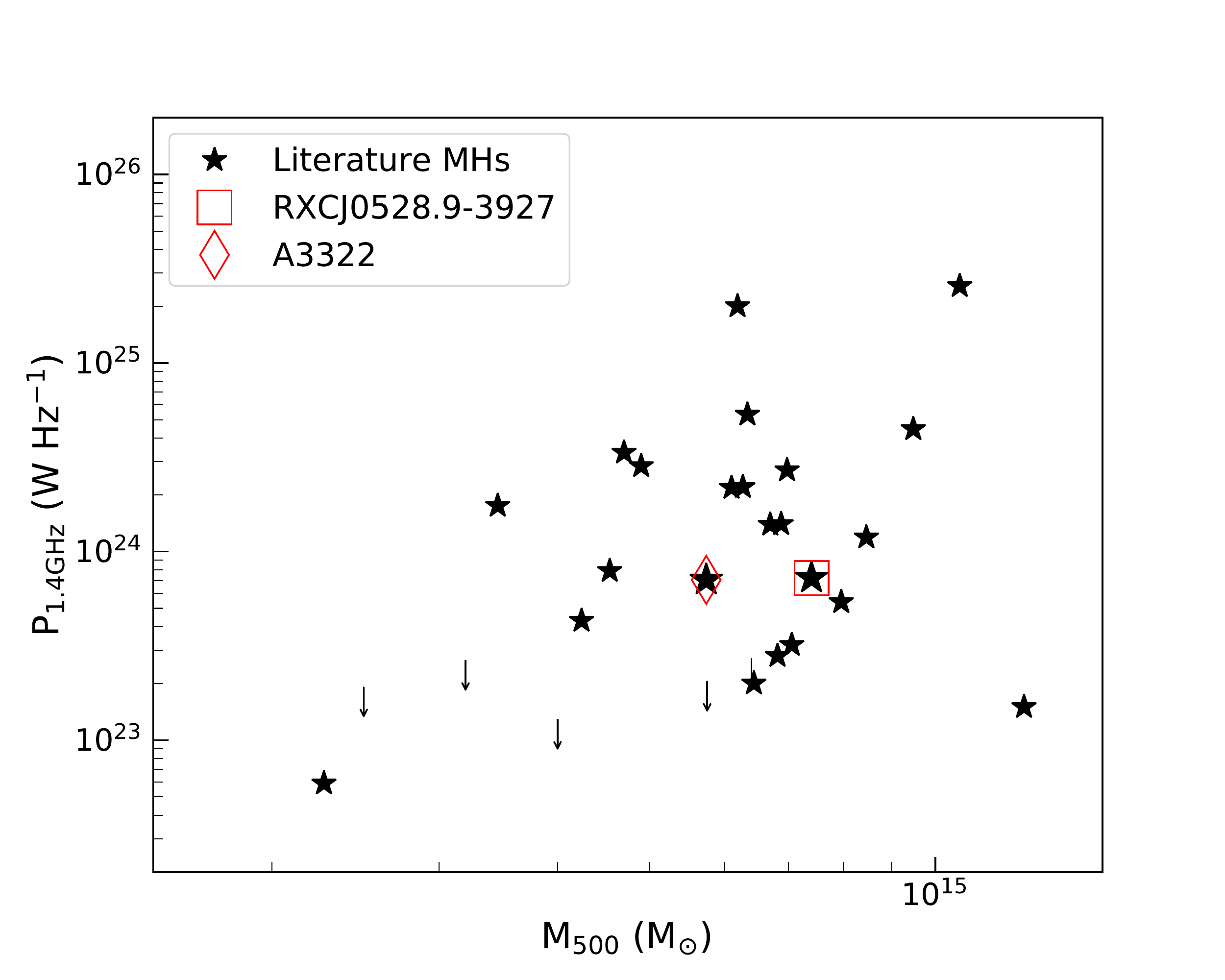}
    \caption{The 1.4 GHz, radio power versus mass plane is shown for clusters hosting radio halos (left) and those hosting mini-halos (right). The arrows show the upper limits for the radio halos and mini-halos from \citet{kal15}. The literature points for radio halos are from \citet{cas13} and for mini-halos are from \citet{gia14}. SUCCESS clusters are highlighted in the plots with open symbols.
    }
    \label{rhmhplot}
\end{figure*}

Scaling relations between radio halo (mini-halo) powers and cluster masses have been studied for obtaining insights on the relation between global cluster properties and the energy in the non-thermal components \citep[e. g.][]{bru07,bas12,ven08,cuciti21b,kal15,laferriere20}. We plot the detections in the SUCCESS clusters in the radio power versus a mass plane (Fig.~\ref{rhmhplot}) along with the sample of radio halos, mini-halos, and upper limits known in the literature \citep[][]{cas13,kal15,gia19}. 
We have updated the position of the radio halo in RXC J0232.2-4420 with the newly determined 1.4 GHz power and mass from \citet{2016A&A...594A..27P} as compared to that in \citet{2019MNRAS.486L..80K}. The radio halo in A3399 and the mini-halo in RXC J0528.9-3927, and the candidate mini-halo in A3322 are also within the range of the scatter in the rest of the sample (Fig.~\ref{rhmhplot}).

\begin{figure}
    \centering
        \includegraphics[height = 6.2cm]{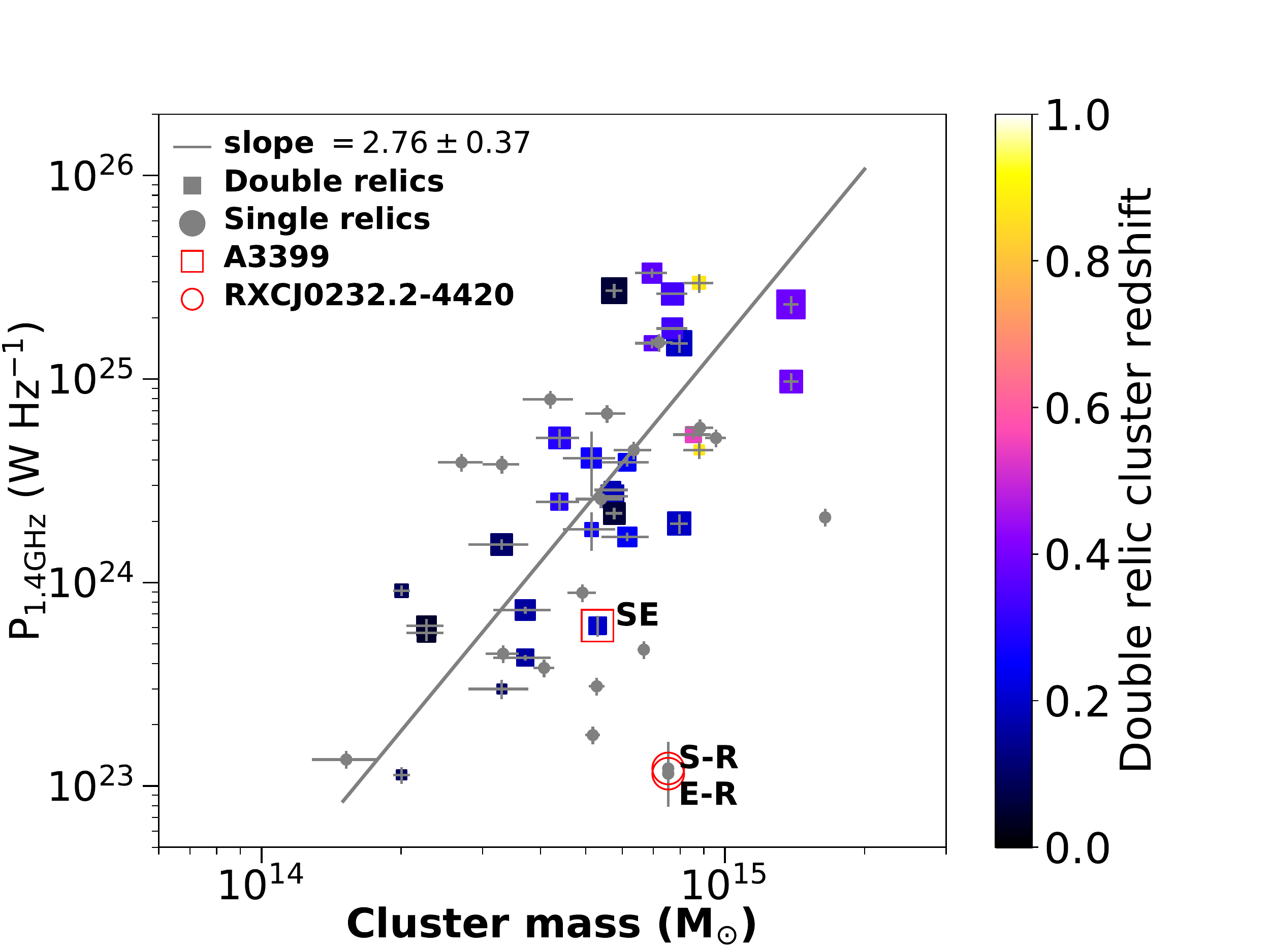}
    \caption{The radio power-mass plane is shown for radio relics with the double relic host cluster redshifts shown in colour and symbol size scaled with the largest linear size of the relic. The SE-relic in A3399 is shown (labeled SE). The flux density of the NW relic could not be unambiguously measured and hence is not shown in the plot. The candidate relics in the RXC J0232.2-4420 (labeled S-R and E-R) are also shown. These being outliers is an indication that these are not associated with the cluster (see Sec.~\ref{sec:rxcj0232}). The literature sample of radio relics is from \citet{2017MNRAS.472..940K}, \citep{2018MNRAS.477..957D} and \citep{2020MNRAS.499..404P}.}
    \label{relicmassplot}
\end{figure}

Radio relics are also known to show a correlation between radio power and mass \citep{gas14,2017MNRAS.472..940K}. We plotted the SE radio relic found in A3399 and the candidate relics in the RXC J0232.2-4420 along with the known sample of relics (Fig.~\ref{relicmassplot}). The flux density of the NW relic in A3399 could not be constrained from the available images, and thus only the SE relic is plotted. This cluster adds to the sample nearby (z$<0.4$) double relic systems, and the dearth of known double relic systems at higher redshifts is noticed in the plot. The candidate relics near RXC J0232.2-4420 are outliers, possibly indicating that they are not related to this cluster.

\subsection{Dynamical states of the clusters}
The dynamical properties of galaxy clusters can be studied with the parameters quantifying their X-ray morphology \citep[e. g.][]{2008A&A...483...35S,2013A&A...549A..19W,2013AstRv...8a..40R,2015A&A...575A.127P,2017ApJ...846...51L}. Typically concentration, centroid shift and power ratios are used. Disturbed or non-relaxed clusters are well separated from their counterparts of relaxed or cool-core clusters on the morphology parameter planes and have been used to establish the role of the dynamical state and the diffuse radio emission \citep[e. g.][]{cas10}. {In order to obtain the dynamical state of our sample clusters, we used the recent work of \cite{2017ApJ...846...51L} in which they computed seven different parameters to classify the X-ray morphologies. A combination of parameters was used to finally assign one of the relaxed, mixed, and disturbed categories to a cluster. While the relaxed and disturbed clusters show clear signatures, the mixed category includes the ones that are intermediate to the two extremes. 
}

\begin{figure}
    \centering
        \includegraphics[trim = 0cm 0cm 0cm 0cm,clip,height = 6.5cm]{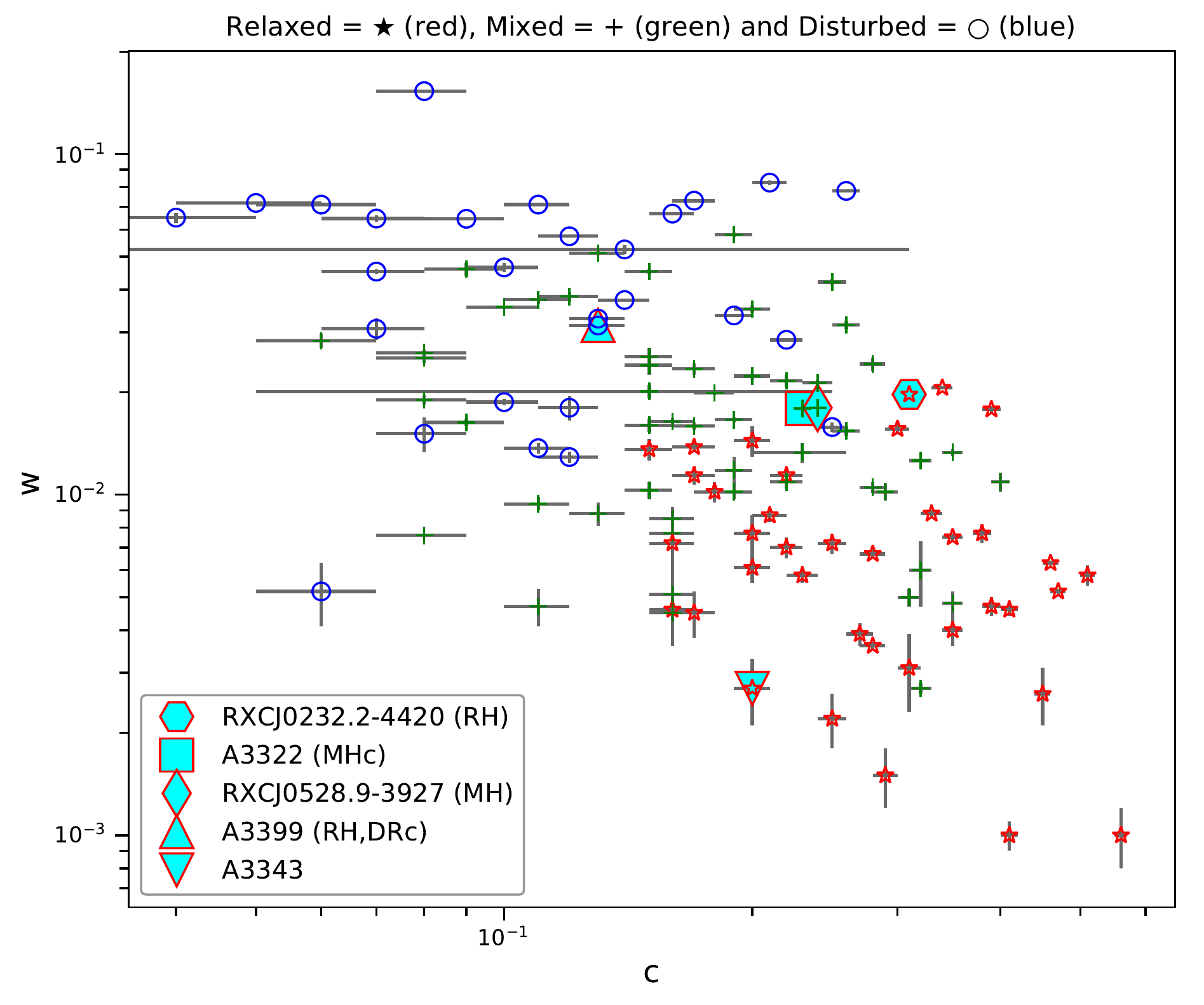}
    \caption{SUCCESS clusters in the c-w plane adapted from \citet{2017ApJ...846...51L}. The red (\textcolor{red}{$\bigstar$}), green (\textcolor{green}{$+$}), and blue (\textcolor{blue}{$\bigcirc$}) indicate relaxed, intermediate, and disturbed clusters, respectively.
    Clusters in our sample are highlighted in the legend. 
    The type of diffuse emission (as listed in Table~\ref{obsn}) is also given in the legend. }
    \label{c-wplot}
\end{figure}

Out of the SUCCESS sub-sample of nine clusters, we found the classification of clusters RXC J0232.2-4420, A3322, RXC J0528.9-3927, A3399, and A3343 in \citet{2017ApJ...846...51L}. 
We show their position with respect to the rest of the sample of clusters in the plane of centroid shift ($w$) versus concentration parameter ($c$) in Fig. \ref{c-wplot}.
Based on this classification, RXC J0232.2-4420, host to a moderate-sized radio halo, is a relaxed cluster and could be a rare system undergoing a transition from mini-halo to radio halo \citet{2019MNRAS.486L..80K}. This is thus among the rare class of systems with relaxed morphology hosting radio halos \citep[e. g.][]{bon14,kalpar16}. A detailed substructure analysis of this cluster by \citet{2021MNRAS.504..610P} using the Chandra and XMM Newton data has shown that the core of the cluster is undisturbed, but a sub-structure to the southwest may be injecting turbulence.
{A3322 and RXC J0528.9-3927 are in mixed (intermediate to relaxed and merging) dynamical states.} The cluster A3399, where double relics and a halo are discovered, is in the disturbed cluster category and is consistent with the dynamical state of all the known clusters hosting double radio relics. It is a rare system as a number of other double relics have been found to be devoid of radio halos \citep{bon17}.

For the four remaining clusters, we have looked for information on the dynamical states in other works.
\citet{2015PASJ...67...71K} have reported that RXC J1358.9-4750 is a pre-merger cluster showing an X-ray bridge between the two clusters. We inspected the XMM-Newton archival image of A3396 and found no obvious sign of disturbance or departure from symmetry and thus can be considered relaxed. There are no high-resolution X-ray data available for PSZ1 G313.85+19.21 except for low-resolution ROSAT data. The ROSAT image suggests that it is an elongated cluster in the northwest to southeast direction and has a total extent of 3.5$'$. It is difficult to state the dynamical state of this cluster due to the limited resolution of these data. We did not find X-ray morphological information for A3937. 

Besides the nine clusters presented in this work, there are 11 clusters in the SUCCESS sample (Table~\ref{clusinfo}). Of these 11, the morphological types of four more are available in \citep{2017ApJ...846...51L}. Among these A3378, A3364 and RXC J0532.9-3701 are classified as relaxed. A3888, the cluster where a radio halo is known, has a mixed morphology. A thorough radio investigation of the SUCCESS sample in radio bands is needed to find any trends between the morphological type and diffuse radio emission.

\section{Summary and Conclusions}\label{conclusions}
%\newpage
The diffuse radio emission from the ICM is an important probe of the magnetic field and cosmic rays in galaxy clusters. We presented the SoUthern Cluster-sCale Extended Source Survey (SUCCESS) sample of 20 galaxy clusters that was selected to be in the southern declinations ($-50 < \delta < -30$) and to have redshifts $<0.3$ and $\mathrm{M} > 5 \times 10^{14} \mathrm{M}_{\odot}$ in the Planck and SPT catalogues.
In this work, we presented radio band studies using targeted GMRT observations of nine clusters in the SUCCESS sample. We also used MeerKAT survey images from the data release 1 of MGCLS survey \citep{2022A&A...657A..56K} for five of these clusters in this study.
The detections and non-detections of diffuse radio emission were reported, and the nature of the diffuse radio sources was interpreted considering the information on the dynamical states of the clusters as reported in the literature. In the following points, we summarize our results and conclusions:
\begin{itemize}
    \item We have presented new GMRT images at 610 MHz for the clusters PSZ1 G313.85+19.21, RXC J0528.9-3927, A3322, and A3937 and at 325 MHz for A3396 and RXC J1358.9-4750. The GMRT 610 MHz results towards RXC J0232.2-4420 were presented in \citet{2019MNRAS.486L..80K}. We have used MeerKAT 1283 MHz images from MGCLS for the clusters RXC J0232.2-4420, RXC J0528.9-3927, A3322, A3399, and A3343.
    \item {We report the MeerKAT detection of the known radio halo in RXC J0232.2-4420 and find the spectral index, $\alpha_{610}^{1283} = 1.6\pm0.4$. In addition, we find two candidate radio relics at distances 1 and 1.9 Mpc from the cluster center. The optical images do not show an obvious, unique identification of any galaxy with the relics.}
    \item We report the GMRT detection of a radio mini-halo in RXC J0528.9-3927 and a candidate mini-halo in A3322 with largest extents of 300 and 248 kpc, respectively. 
    \item From the GMRT and MeerKAT images, catalogs of discrete radio sources within the extents of the detected diffuse emission are presented for the clusters RXC J0528.9-3927, A3399, A3322, and RXC J0232.2-4420.
    \item {We further examined the locations of the detected radio halos, mini-halos and relics in the radio power versus host cluster mass plane. The radio halos and mini-halos show a behavior consistent with the earlier sample known from the literature. The southeast relic in A3399 is consistent with the radio relic radio power-mass scaling. The candidate relics around RXC J0232.2-4420 are outliers; this further indicates that they are likely of an origin different from cluster merger.}
    \item The dynamical states of five SUCCESS clusters were found in the sample studied by \citet{2017ApJ...846...51L}. RXC J0232.2-4420 (RH) and non-detection A3343 are relaxed, A3399 (RH+DRc) is disturbed, and A3322 (MHc) and RXCJ0528.9-3937 (MH) show mixed morphology. 
    The relaxed morphology of RXC J0232.2-4420 (RH) is a surprise, and this has been discussed 
    as a case of a mini-halo to halo transition by \citet{2019MNRAS.486L..80K}.
    \item We found that within the SUCCESS sub-sample of 9 clusters, $11\%$  clusters contain radio relics, $22\%$ clusters contain radio halos, and the same fraction contains mini-halos. {We note that the sample has only 9 clusters and that the non-detections are likely due to the sensitivity limitation of the GMRT in the cases of A3396 and PSZ2G313.84+1921.} We also note that A3396 and A3343 are low mass clusters ($<5\times10^10^{14}$ M$_\odot$) where the chance of occurrence of radio halos is low.
    \item The SUCCESS sample represents nearby, massive clusters in the southern part of the sky which was less explored until recently. 
    In future we aim to complete the radio survey of SUCCESS sample with the Upgraded GMRT and MeerKAT in order to carry out a statistical study of occurrence of diffuse radio sources in nearby clusters.
\end{itemize}

\section*{Acknowledgements}
We thank the referee for their comments that have improved the clarity of the paper. 
R. K. thanks A. Botteon for providing the X-ray image of the cluster RXC J0528.9-3927. R.K. acknowledges the support of the Department of Atomic Energy, Government of India, under project no. 12-R\&D-TFR-5.02-0700. This research has made use of the data from {\it GMRT} Archive. 
This research has made use of data obtained through the High Energy Astrophysics Science Archive Research Center Online Service, provided by the NASA/Goddard Space Flight Center.
V. P. acknowledges the South African Research Chairs Initiative of the Department of Science and Technology and National Research Foundation who supported this research. V.P. also acknowledges the financial assistance of the South African Radio Astronomy Observatory (SARAO) towards this research. 
M. R. acknowledges the support provided by the DST-Inspire fellowship program (IF160343) by DST, India. M. R. also acknowledges support from Ministry of Science and Technology of Taiwan (MOST 109-2112-M-007-037-MY3).
TV acknowledges the support from the Ministero degli Affari Esteri edella Cooperazione Internazionale, Direzione Generale per la Promozionedel Sistema Paese, Progetto di Grande Rilevanza ZA18GR02.
K.K. is supported in part by the National Research Foundation of South Africa (N.R.F. Grant Numbers: 120850). Opinions, findings, conclusions, or recommendations expressed in this publication are that of the author(s), and the N.R.F. accepts no liability whatsoever in this regard.
S. P. S. acknowledges funding support from N.R.F./South African Radio Astronomy Observatory.
D. S. P. acknowledges funding from the South African Radio Astronomy Observatory and the National Research Foundation (N.R.F. Grant Number: 96741).
We thank the staff of the GMRT that made these observations possible. GMRT is run by the National Centre for Radio Astrophysics of the Tata Institute of Fundamental Research. The MeerKAT telescope is operated by the South African Radio Astronomy Observatory, which is a facility of the National Research Foundation, an agency of the Department of Science and Innovation. This research has made use of  NASA's  Astrophysics Data  System and of the NASA/IPAC Extragalactic Database  (N.E.D.), which is operated by the Jet  Propulsion Laboratory, California Institute of Technology, under contract with the National Aeronautics and Space Administration.
%Facilities: Chandra (GMRT), SDSS.

%The Acknowledgements section is not numbered. Here you can thank helpful
%colleagues, acknowledge funding agencies, telescopes and facilities used etc.

%%%%%%%%%%%%%%%%%%%%%%%%%%%%%%%%%%%%%%%%%%%%%%%%%%
\section*{Data Availability}
The data underlying this article will be shared on reasonable request to the corresponding author.


\bibliographystyle{mnras}
\bibliography{ruta_all_1,VP_references,majidul} % if your bibtex file is called example.bib
%\bibliography{VP_references}

%\begin{thebibliography}{99}
%\end{thebibliography}

%%%%%%%%%%%%%%%%%%%%%%%%%%%%%%%%%%%%%%%%%%%%%%%%%%

%%%%%%%%%%%%%%%%% APPENDICES %%%%%%%%%%%%%%%%%%%%%
\appendix

\section{Discrete and diffuse sources in MeerKAT images}\label{app:meerkat}
The discrete sources in MeerKAT images and the regions used to find diffuse emission flux densities are marked on the MeerKAT images and shown here for the 4 clusters with diffuse radio emission.
\begin{figure}
\centering
\includegraphics[width=\columnwidth]{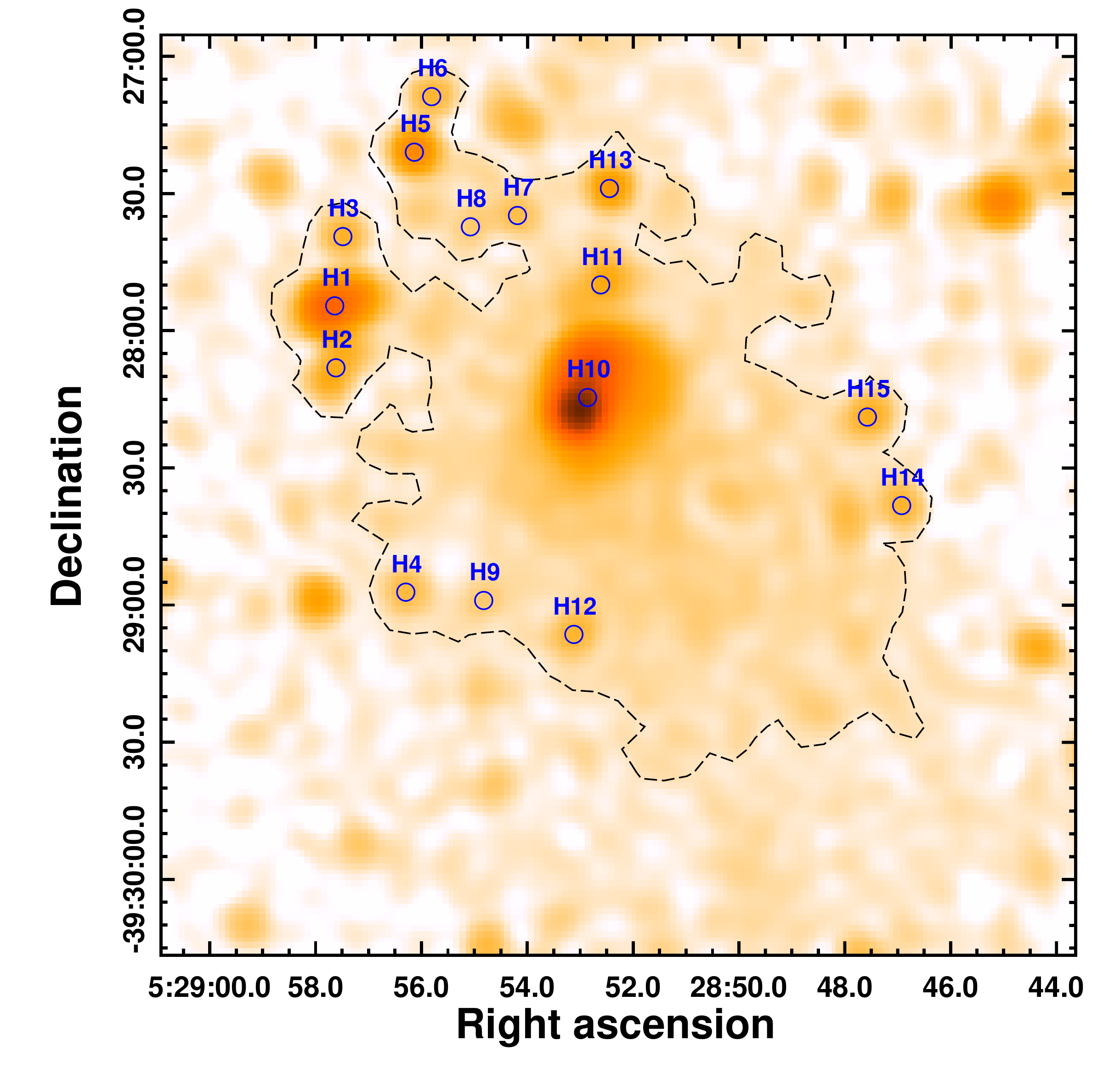} \\
\caption{\textbf{RXC J0528.9-3927}: MeerKAT 1283 MHz image with a resolution of $7.3''\times7.3''$ 
and rms, $\sigma_{\mathrm{rms}} = 2.6 \mu$Jy beam$^{-1}$  of the cluster zoomed in on the diffuse source is shown. The region of the diffuse emission ($>3\sigma_{\mathrm{rms}}$, dashed line) and the labels of the discrete sources within this region found using PyBDSF are marked. 
}
\label{appfig:rxcj0528}
\end{figure}

\begin{figure*}
\centering
\includegraphics[width=6.5in]{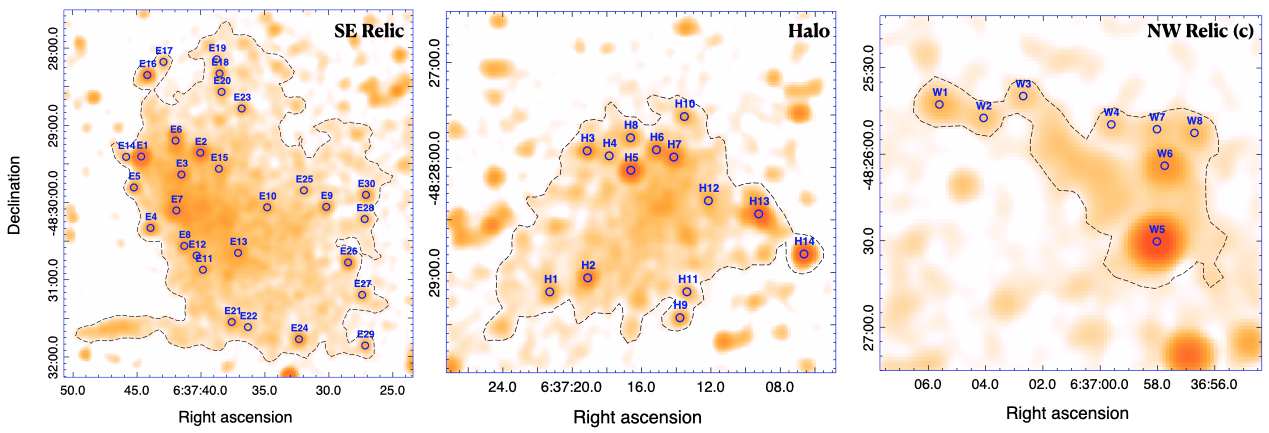} \\
\caption{\textbf{A3399}: MeerKAT 1283 MHz image with a resolution of $7.5''\times 7.3''$ 
and rms, $\sigma_{\mathrm{rms}} = 2.9 \mu$Jy beam$^{-1}$  of the cluster zoomed in on the diffuse sources are shown in the three panels. 
The region of the diffuse emission ($> 3 \sigma_{\mathrm{rms}}$, dashed line) and the labels of the discrete sources within this region found using PyBDSF are marked. Left: Region of the SE relic. Middle: Region of the radio halo. Right: Region of the NW relic.
}
\label{appfig:a3399}
\end{figure*}

\begin{figure}
\centering
\includegraphics[width=\columnwidth]{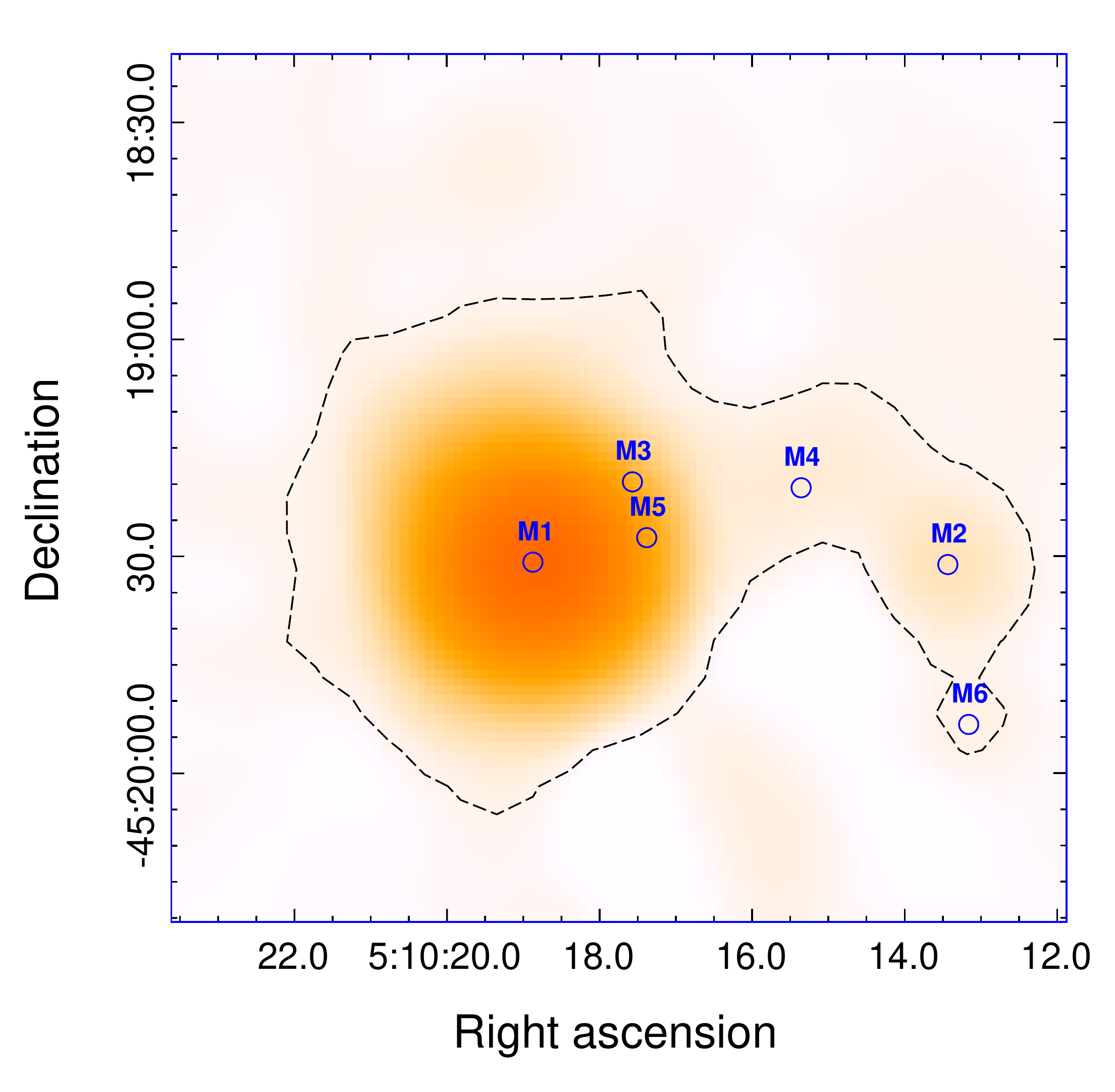} \\
\caption{\textbf{A3322}: MeerKAT 1283 MHz images with resolutions of $15''\times15''$  and $\sigma_{\mathrm{rms}} = 15.0 \mu$Jy beam$^{-1}$  of the cluster zoomed in on the diffuse source is shown. The region of the diffuse emission ($>3\sigma_{\mathrm{rms}}$, dashed line) and the labels of the discrete sources within this region found using PyBDSF are marked.
}
\label{appfig:a3322}
\end{figure}

\begin{figure*}
\centering
\includegraphics[width=6.5in]{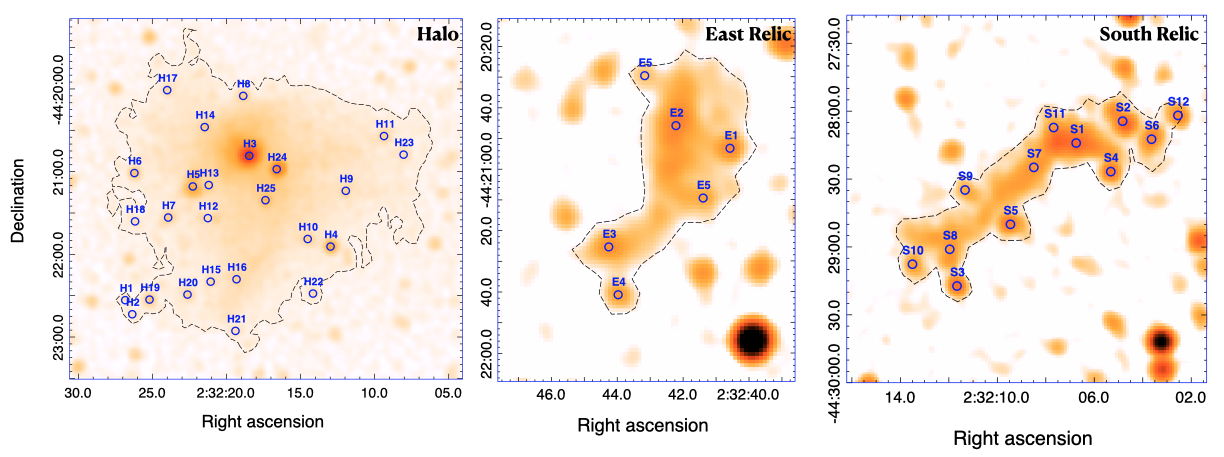}
\caption{\textbf{RXC J0232.2-4420}: The MeerKAT 1283 MHz image with a resolution of $7.1''\times7.1''$ and rms, $\sigma_{\mathrm{rms}} = 2.6\mu$Jy beam$^{-1}$ of the cluster zoomed in on the diffuse sources is shown in the three panels. The region of the diffuse emission ($>3\sigma_{\mathrm{rms}}$, dashed line) and the labels of the discrete sources within this region found using PyBDSF are marked. Left: Region of the radio halo. Middle: Region of the East relic. Right: Region of the southern relic.}
\label{appfig:rxcj0232}
\end{figure*}


%%%%%%%%%%%%%%%%%%%%%%%%%%%%%%%%%%%%%%%%%%%%%%%%%%

% Don't change these lines
\bsp	% typesetting comment
\label{lastpage}
\end{document}